\documentclass[runningheads]{llncs}

\usepackage{amssymb}
\usepackage{caption}
\usepackage{subcaption}
\usepackage[title]{appendix}

\usepackage{balance}		
\usepackage{graphics}		
\usepackage{url}			
\usepackage{dblfloatfix}	
\usepackage{stackengine}
\usepackage{pbox}
\usepackage{array}

\usepackage[utf8]{inputenc}
\usepackage{graphicx}
\usepackage{colortbl,array} 
\usepackage{multirow,bigdelim}
\usepackage{booktabs}
\usepackage{listings}
\usepackage{color}
\usepackage{colortbl,array} 
\usepackage{mathtools}
\usepackage{blkarray, bigstrut} %
\usepackage{amsmath,kbordermatrix,blkarray}
\usepackage{algorithm}
\usepackage[noend]{algpseudocode}
\usepackage{todonotes}

\usepackage{tikz}

\usetikzlibrary{calc}

\usepackage{verbatim}
\usetikzlibrary{arrows,shapes}

\renewcommand{\kbldelim}{[}
\renewcommand{\kbrdelim}{]}

\usepackage{tikz-cd}
\usetikzlibrary{matrix,arrows}
\usetikzlibrary{positioning,calc}

\urldef{\mailsa}\path|{roblim1,norris,malony}@cs.uoregon.edu|

\usepackage{colortbl,array} 

\newcommand\tabhead[1]{\small\textbf{#1}}

\newcolumntype{P}[1]{>{\centering\arraybackslash}p{#1}}
\newcolumntype{M}[1]{>{\centering\arraybackslash}m{#1}}

\definecolor{mygreen}{rgb}{0,0.6,0}
\definecolor{mygray}{rgb}{0.5,0.5,0.5}
\definecolor{mymauve}{rgb}{0.58,0,0.82}
\definecolor{myltgray}{rgb}{0.83,0.83,0.83}

\usepackage{adjustbox}\usepackage{mathtools}
    \usepackage{array}
    \usepackage{makecell}
    \usepackage{stackengine}
\setstackEOL{\cr} 

\begin{document}

\mainmatter  

\title{A Similarity Measure for GPU Kernel\\Subgraph Matching}
\titlerunning{A Similarity Measure for GPU Kernel\\Subgraph Matching}

\author{Robert Lim%
\and Boyana Norris 
\and Allen Malony}
\authorrunning{Lim, Norris, Malony}

\institute{
University of Oregon, Eugene, OR, USA\\
\mailsa\\
}

%
%

\toctitle{Lecture Notes in Computer Science}
\tocauthor{Authors' Instructions}
\maketitle

\begin{abstract}
Accelerator architectures specialize in executing SIMD (single instruction, multiple data) in lockstep.  Because the majority of CUDA applications are parallelized loops, control flow information can provide an in-depth characterization of a kernel.  \texttt{CUDAflow} is a tool that statically separates CUDA binaries into basic block regions and dynamically measures instruction and basic block frequencies.  \texttt{CUDAflow} captures this information in a control flow graph (CFG) and performs subgraph matching across various kernel's CFGs to gain insights into an application's resource requirements, based on the shape and traversal of the graph, instruction operations executed and registers allocated, among other information.  The utility of \texttt{CUDAflow} is demonstrated with SHOC and Rodinia application case studies on a variety of GPU architectures, revealing novel control flow characteristics that facilitate end users, autotuners, and compilers in generating high performing code.
\end{abstract}

\section{Introduction}

Structured programming consists of base constructs that represent how
programs are written \cite{bohm1966flow,williams1978conversion}.  When
optimizing programs, compilers typically operate on the intermediate
representation (IR) of a control flow graph (CFG), which is derived from
program source code analysis and represents basic blocks of instructions
(nodes) and control flow paths (edges) in the graph.  Thus, the overall
program structure is captured in the CFG and the IR abstracts
machine-specific intrinsics that the compiler ultimately translates to
machine code.  The IR/CFG allows the compiler to reason more efficiently 
about optimization opportunities and apply transformations.  In particular,
compilers can benefit from prior knowledge of optimizations that may be
effective for specific CFG structures.

In the case of accelerated architectures that are programmed for SIMD
parallelism, control divergence encountered by threads of execution
presents a major challenge for applications because it can severely reduce
SIMD computational efficiency.  It stands to reason that by identifying the
structural patterns of a CFG from an accelerator (SIMD) program, insight on
the branch divergence problem \cite{sabne2016formalizing} might be gained
to help in their optimization.  Current profiling approaches to understanding
thread divergence behavior
(e.g., \cite{ddt,nvprof,shende2006tau})
do not map performance
information to critical execution paths in the CFG.  While accelerator devices (e.g.,
GPUs) offer hardware performance counters for measuring computational
performance, it is more difficult to apply them to capture divergence
behavior \cite{lim2015identifying}.

Our research focuses on improving the detail and accuracy of control flow
graph information in accelerator (GPU) programs.  We study the
extent to which CFG data can provide sufficient context for understanding a
GPU kernel's execution performance.  Furthermore, we want to investigate
how effective knowledge of CFG shapes (patterns) could be in enabling
optimizing compilers and autotuners to infer execution characteristics
without having to resort to running execution experiments.  To this end, we
present \texttt{CUDAflow}, a scalable toolkit for heterogeneous computing
applications.  Specifically, \texttt{CUDAflow} provides a new methodology
for characterizing CUDA kernels using control flow graphs and instruction
operations executed.  It performs novel kernel subgraph matching to gain
insights into an application's resource requirements.  To the knowledge of
the authors, this work is a first attempt at employing subgraph matching
for revealing control flow behavior and generating efficient
code.

Contributions described in this paper include the following.
\begin{itemize}
\item
  Systematic process to construct control flow graphs for GPU kernels.
\item
  Techniques to perform subgraph matching on various kernel CFGs and GPUs.
\item
  Approaches to reveal control flow behavior based on CFG properties.
\end{itemize}

The rest of the paper is organized as follows.  Section~\ref{sec:prior} discusses prior work, and Section~\ref{sec:background} provides background information.  Section~\ref{sec:methodology} describes the methodology behind our \texttt{CUDAflow} tool and our implementation approach.  Sections~\ref{sec:experiments} and~\ref{sec:results} summarizes the findings of our application characterization studies. Section~\ref{sec:conclusions} outlines future work.

\section{Prior Work}
\label{sec:prior}
Control flow divergence in heterogeneous computing applications is a well known and difficult problem, due to the lockstep nature of the GPU execution paradigm.  Current efforts to address branch divergence in GPUs draw from several fields, including profiling techniques in CPUs, and software and hardware architectural support in GPUs.  For instance, Sarkar demonstrated that the overall execution time of a program can be estimated by deriving the variances of basic block regions \cite{sarkar1989determining}.  Control flow graphs for flow and context sensitive profiling were discussed in \cite{ammons1997exploiting,ball1994optimally}, where instrumentation probes were inserted at selected edges in the CFG, which reduced the overall profiling overhead with minimal loss of information.  Hammock graphs were constructed \cite{zhang2004using} that mapped unstructured control flow on a GPU \cite{diamos2011simd,wu2011characterization}.  By creating thread frontiers to identify early thread reconvergence opportunities, dynamic instruction counts were reduced by as much as 633.2\%.


Lynx~\cite{farooqui2012lynx} creates an internal representation of a program based on PTX and then emulates it, which determines the memory, control flow and parallelism of the application.  This work closely resembles ours but differs in that we perform workload characterization on actual hardware during execution.  Other performance measurement tools, such as HPCToolkit~\cite{adhianto2010hpctoolkit} and DynInst~\cite{miller1995paradyn}, provide a way for users to construct control flow graphs from CUDA binaries, but do not analyze the results further.  The MIAMI toolkit \cite{marin2014miami} is an instrumentation framework for studying an application's dynamic instruction mix and control flow but does not support GPUs.  

Subgraph matching has been explored in a variety of contexts.  For instance, the DeltaCon framework matched arbitrary subgraphs based on similarity scores \cite{koutra2013d}, which exploited the properties of the graph (e.g., clique, cycle, star, barbell) to support the graph matching.  Similarly, frequent subgraph mining was performed on molecular fragments for drug discovery \cite{borgelt2002mining}, whereas document clustering was formalized in a graph database context \cite{huan2003efficient}.  The IsoRank authors consider the problem of matching protein-protein interaction networks between distinct species \cite{singh2007pairwise}.  The goal is to leverage knowledge about the proteins from an extensively studied species, such as a mouse, which when combined with a matching between mouse proteins and human proteins can be used to hypothesize about possible functions of proteins in humans.  However, none of these approaches apply frequent subgraph matching for understanding performance behavior of GPU applications.

\tikzstyle{decision} = [diamond, draw, fill=green!20, rounded corners,
    text width=4.5em, text badly centered, node distance=2.5cm, inner sep=0pt]
\tikzstyle{block} = [rectangle, draw, fill=orange!20, 
    text width=5em, text centered, rounded corners, minimum height=4em]
\tikzstyle{line} = [draw, -latex']
\tikzstyle{cloud} = [draw, ellipse,fill=red!20, text centered, text width=4em, minimum width=1cm]
\tikzstyle{user} = [draw, ellipse,fill=red!20, text centered, text width=2em, minimum width=1cm]
\tikzstyle{round} = [draw, circle,fill=green!20,
    minimum height=2em]
\tikzstyle{startstop} = [rectangle, rounded corners, minimum width=3cm, minimum height=1cm,text centered, draw=black, fill=red!30]
\tikzstyle{io} = [trapezium, trapezium left angle=70, trapezium right angle=110, draw=black, rounded corners,fill=blue!20, text centered, text width=8.6em]
\tikzstyle{source} = [trapezium, trapezium left angle=70, trapezium right angle=110, draw=black, rounded corners,fill=blue!20, text centered, text width=4.5em]
\tikzstyle{optimal} = [trapezium, trapezium left angle=70, trapezium right angle=110, draw=black, rounded corners,fill=blue!20, text centered, text width=4em]
\tikzstyle{process} = [rectangle, minimum width=3cm, minimum height=1cm, text centered, text width=3cm, draw=black, rounded corners,fill=orange!20]
\tikzstyle{output} = [rectangle, minimum width=3cm, minimum height=1cm, text centered, text width=3cm, draw=black, rounded corners,fill=blue!20]
\tikzstyle{arrow} = [thick,->,>=stealth]
\begin{figure}
\centering
\resizebox{0.6\textwidth}{!}{%
\begin{tikzpicture}[node distance = 2cm, auto]
    \node [round] (nvcc) {nvcc};
    \node [source, above of=nvcc, node distance=1.2cm] (source) {source.cu};
    \node [user, left of=source, node distance=2cm] (user) {user};    
    \node [block, below of=nvcc, node distance=1.5cm] (cfg) {construct CFG};
    \node [block, below of=cfg, node distance=1.5cm] (pc) {sample PC counter};
    \node [cloud, left of=nvcc, node distance=2.5cm ] (profiler) {CUDAflow profiler};    
    \node [decision, right of=cfg,node distance=3cm] (cudaflow) {CUDAflow analysis};
    \node [io, below of=cudaflow, node distance=3cm] (stop) { basic block counts instruction mixes CFG matching};
    \path [line] (nvcc) -- (cfg);
    \path [line] (user) -- (source);
    \path [line] (source) -- (nvcc);    
    \path [line] (cfg) -- (cudaflow);
    \path [line] (pc) -- (cudaflow);
    \path [line,dashed] (cudaflow) -- node {}(stop);
    \path [line,dashed] (profiler) |- (cfg);
    \path [line,dashed] (profiler) |- (pc);
\end{tikzpicture}       
}
\caption{Overview of our proposed \texttt{CUDAflow} methodology.}
\label{fig:flow}
\end{figure}
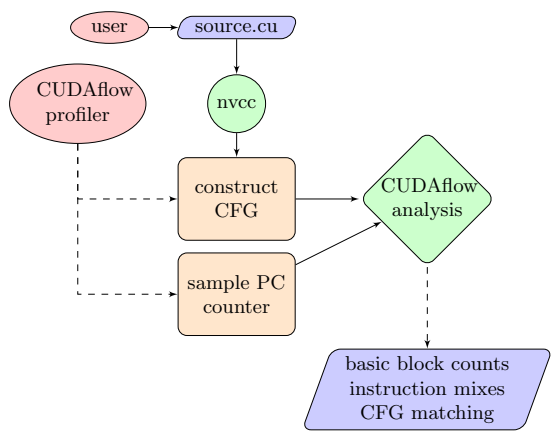

\begin{figure}[thb]
\centering
\vspace{-.1in}
\begin{tabular}
{|M{1.2cm}M{1.2cm}M{1.2cm}|M{1.2cm}M{1.2cm}M{1.2cm}|M{1.2cm}M{1.2cm}M{1.2cm}|}\hline
 \text{\scriptsize Kepler}&\text{\scriptsize Maxwell}&\text{\scriptsize Pascal}&\text{\scriptsize Kepler}&\text{\scriptsize Maxwell}&\text{\scriptsize Pascal} &\text{\scriptsize Kepler} &\text{\scriptsize Maxwell} &\text{\scriptsize Pascal} \\ \hline
\includegraphics[scale=.15]{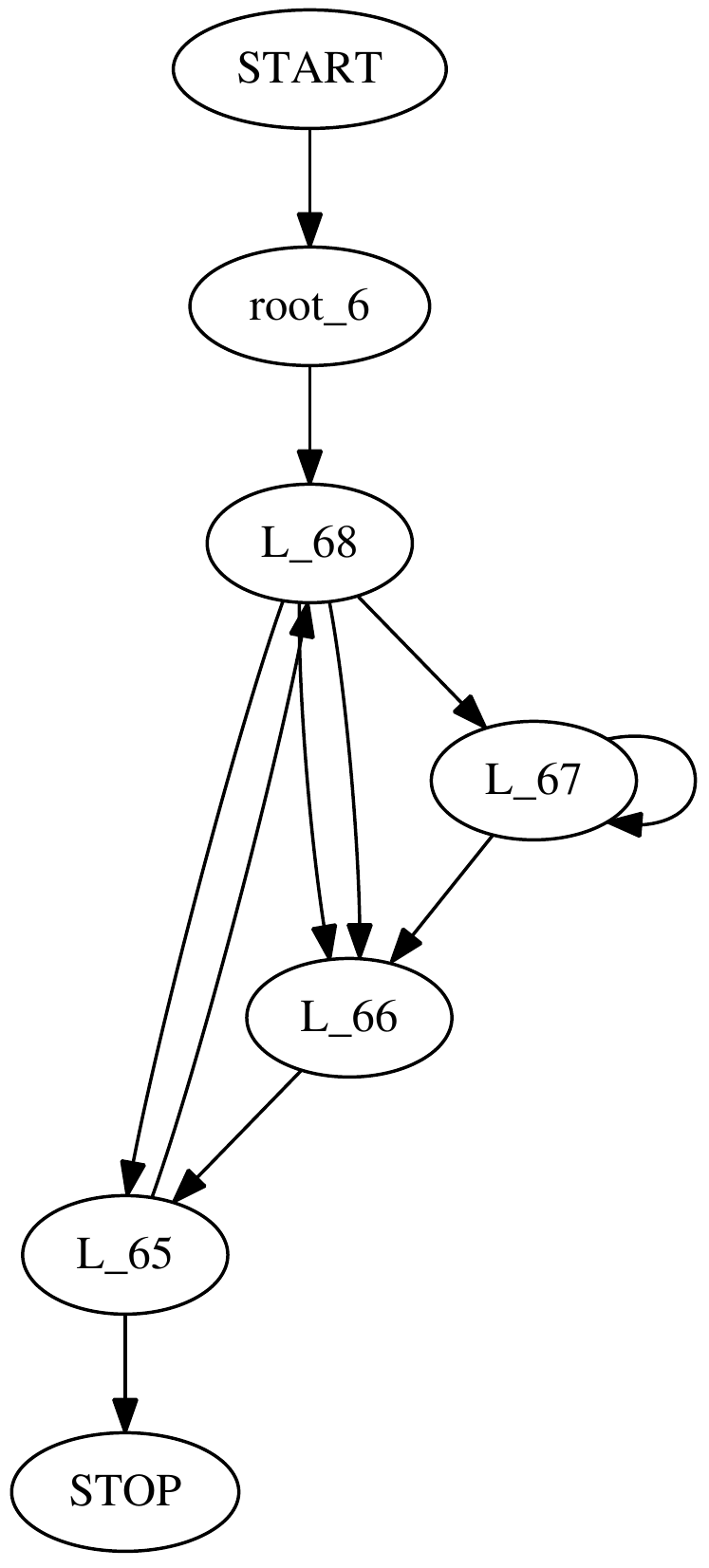} & \includegraphics[scale=.2]{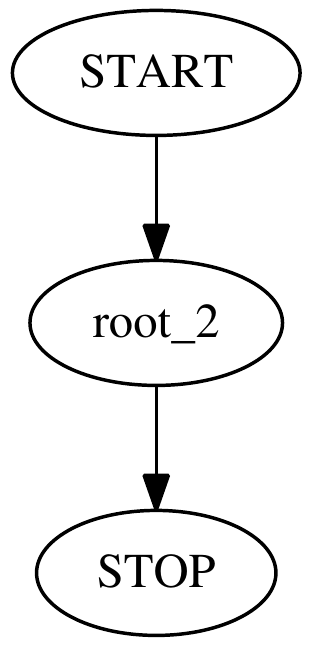} & \includegraphics[scale=.15]{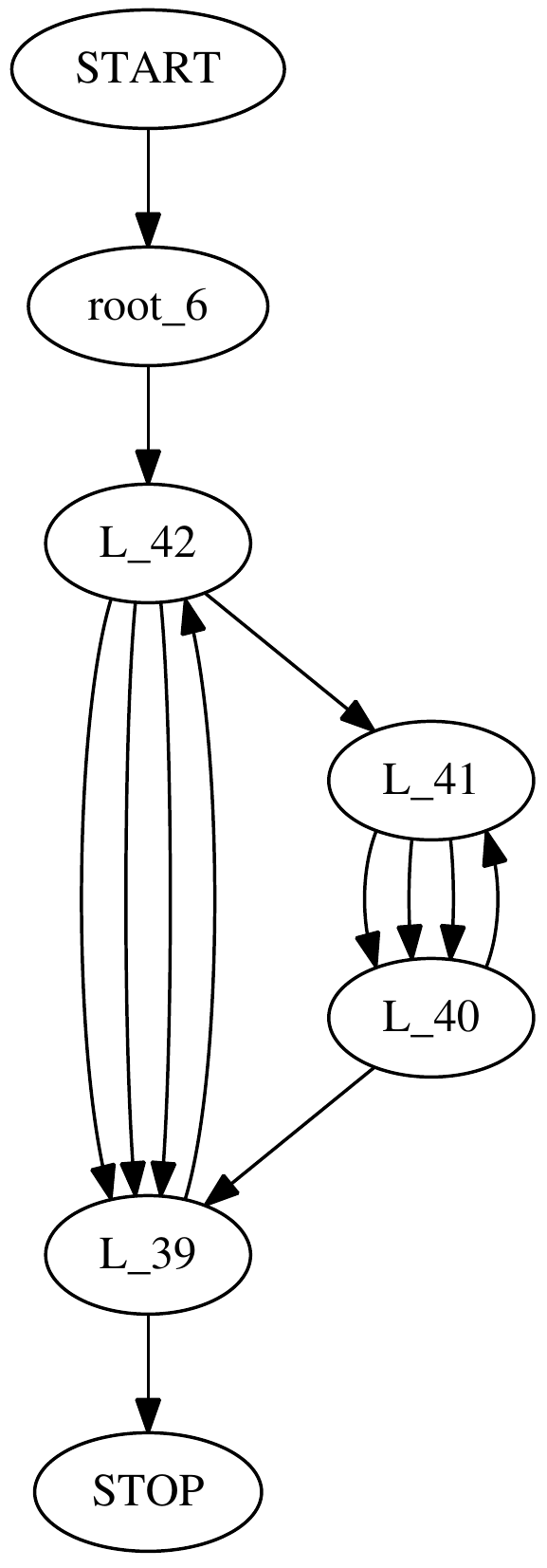}  & \includegraphics[scale=.11]{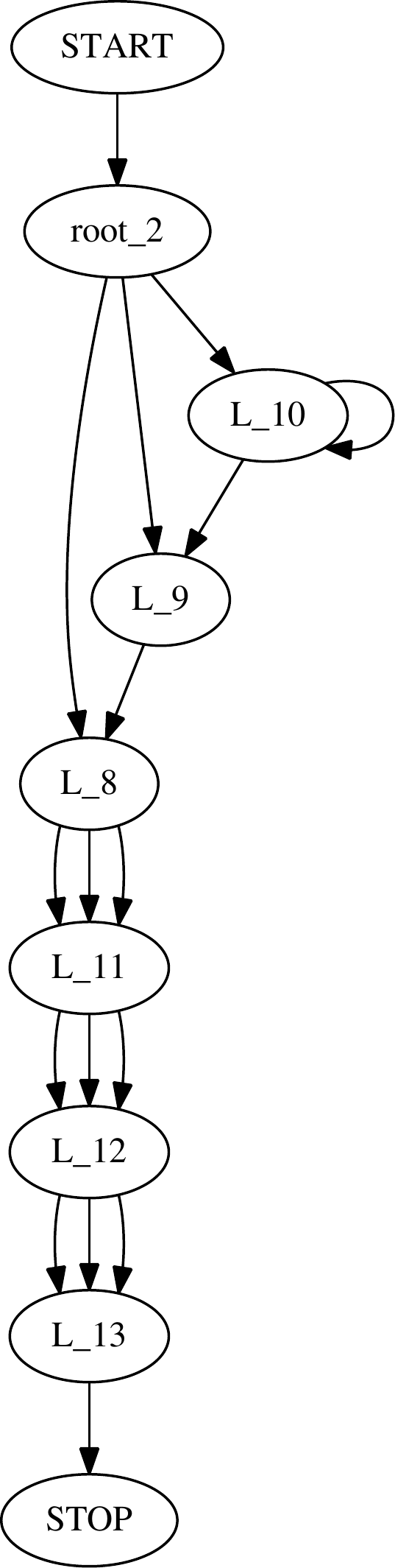} & \includegraphics[scale=.18]{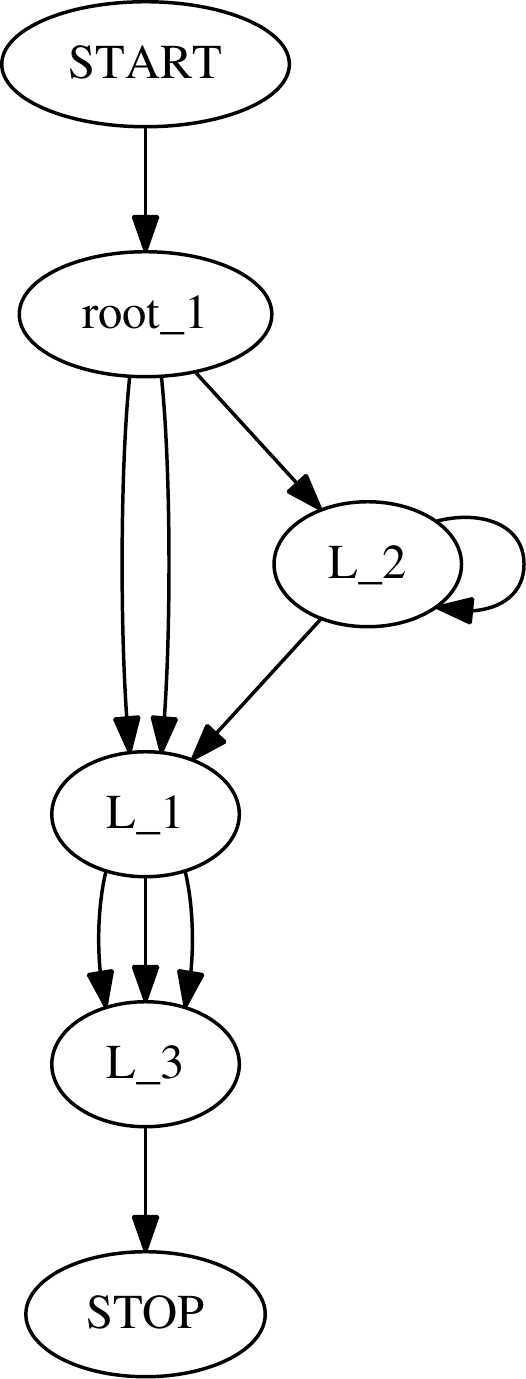} & \includegraphics[scale=.15]{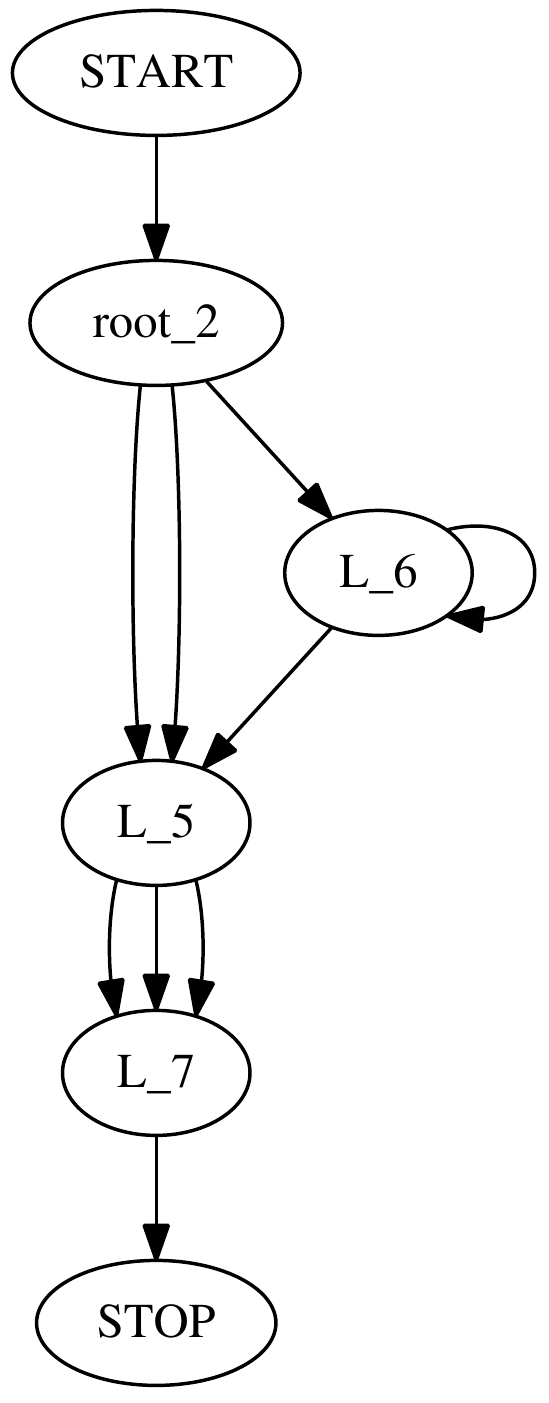} & \includegraphics[scale=.12]{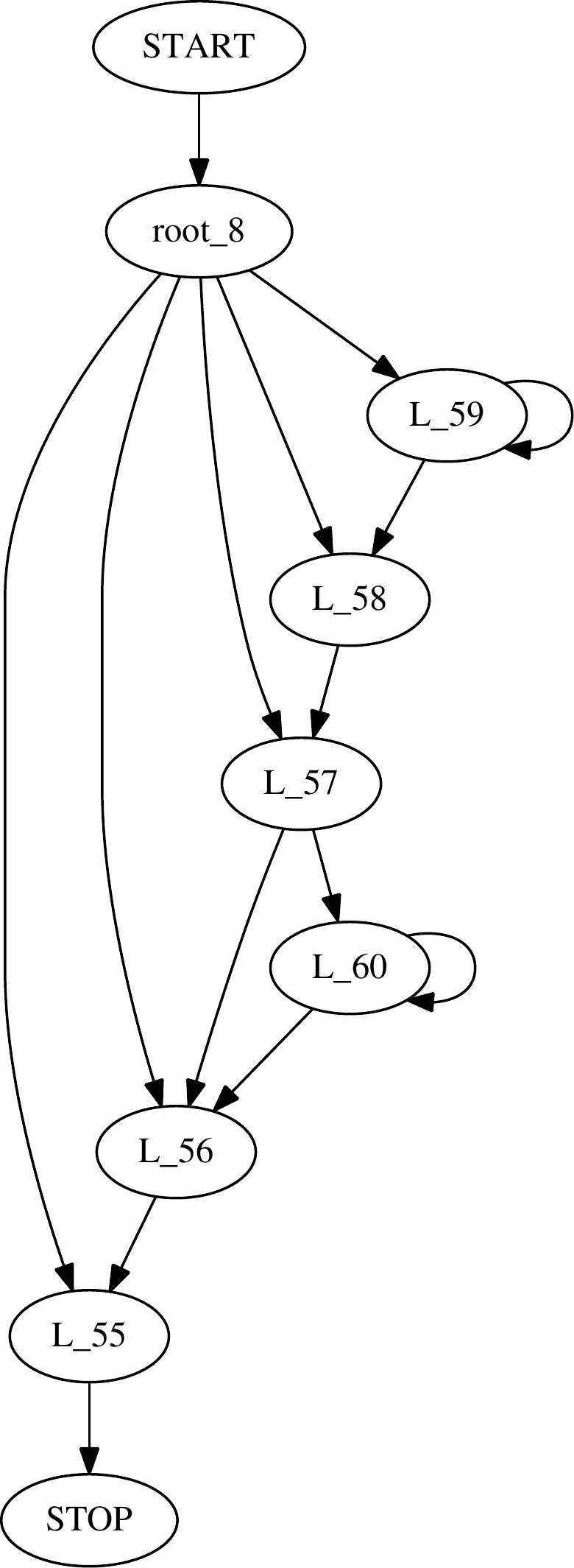} & \includegraphics[scale=.14]{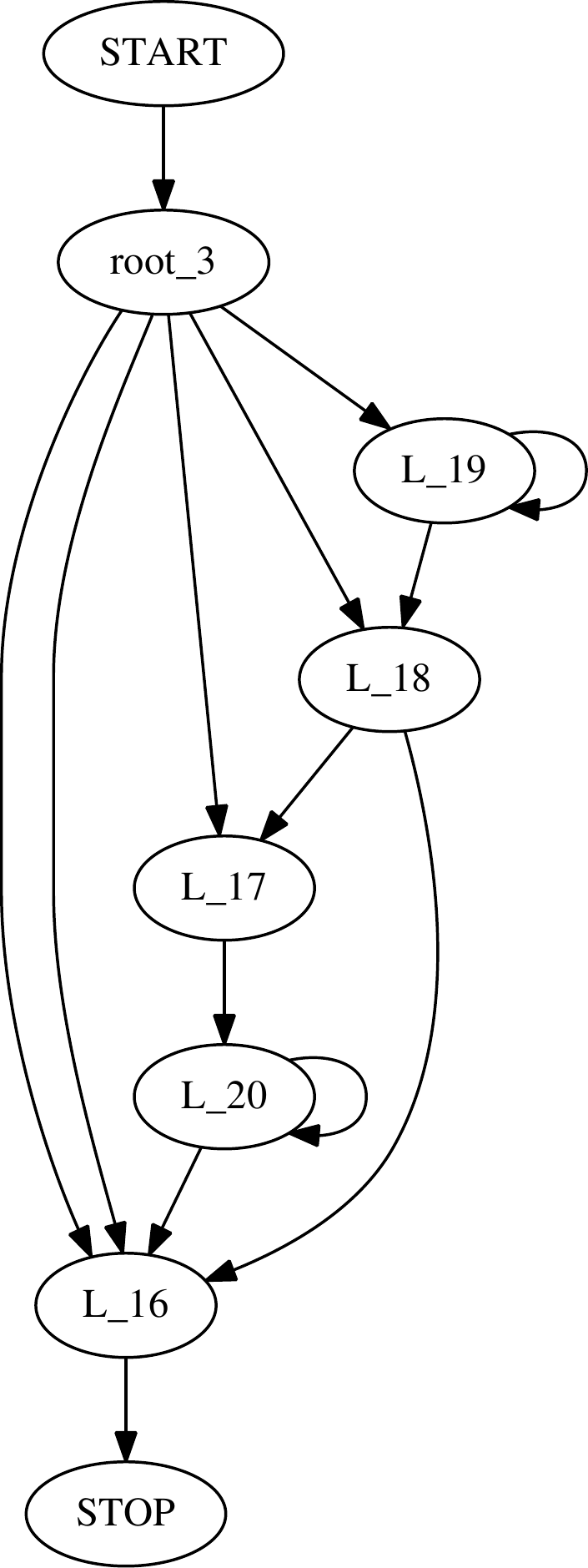} & \includegraphics[scale=.11]{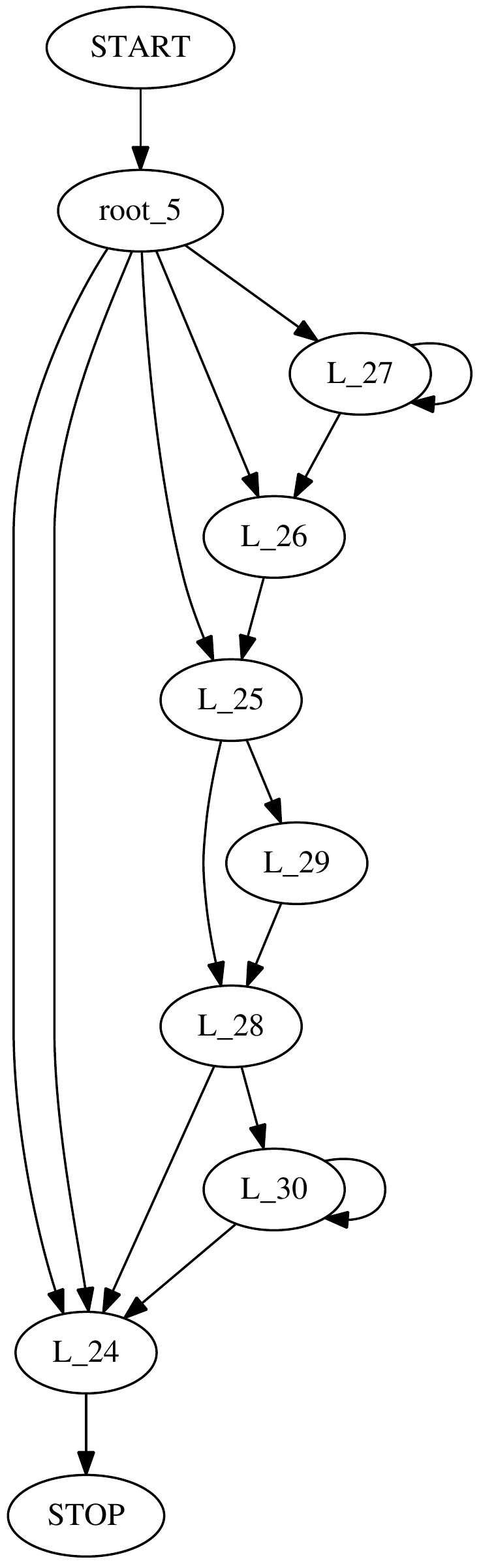} 
\\ \hline
\multicolumn{3}{|p{0.3\columnwidth}|}{\centering\tabhead{\begin{scriptsize}
 BFS \texttt{kernel\_warp}\end{scriptsize}}} &\multicolumn{3}{|p{0.3\columnwidth}|}{\centering\tabhead{ \begin{scriptsize} Reduction \texttt{reduce} \end{scriptsize} }} &\multicolumn{3}{|p{0.3\columnwidth}|}{\centering\tabhead{ \begin{scriptsize}SPMV \texttt{csr\_scalar} \end{scriptsize}}} \\ \hline
\includegraphics[scale=.2]{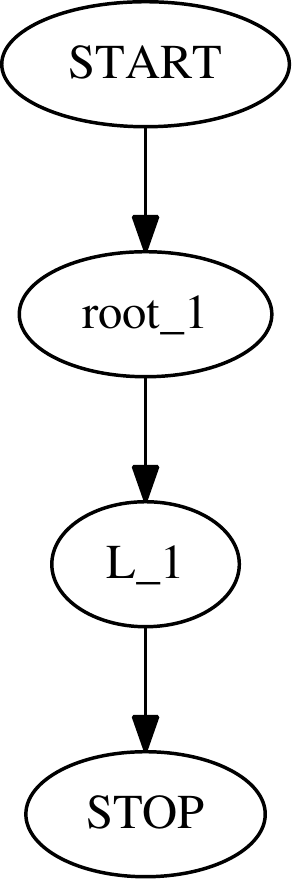} & \includegraphics[scale=.2]{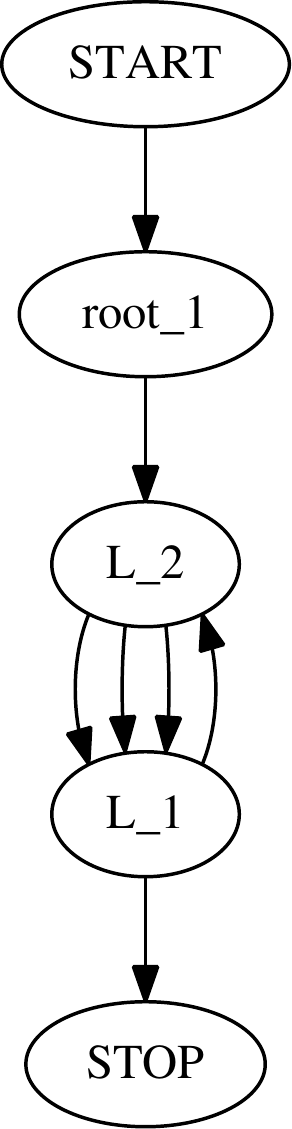} & \includegraphics[scale=.2]{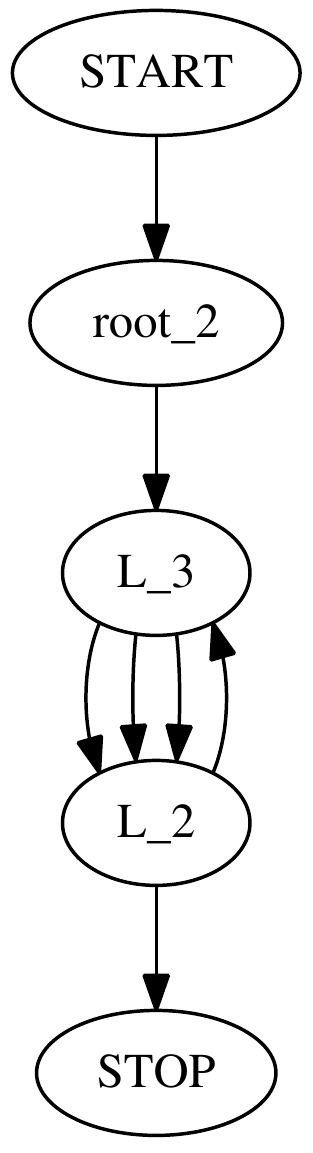} & \includegraphics[scale=.18]{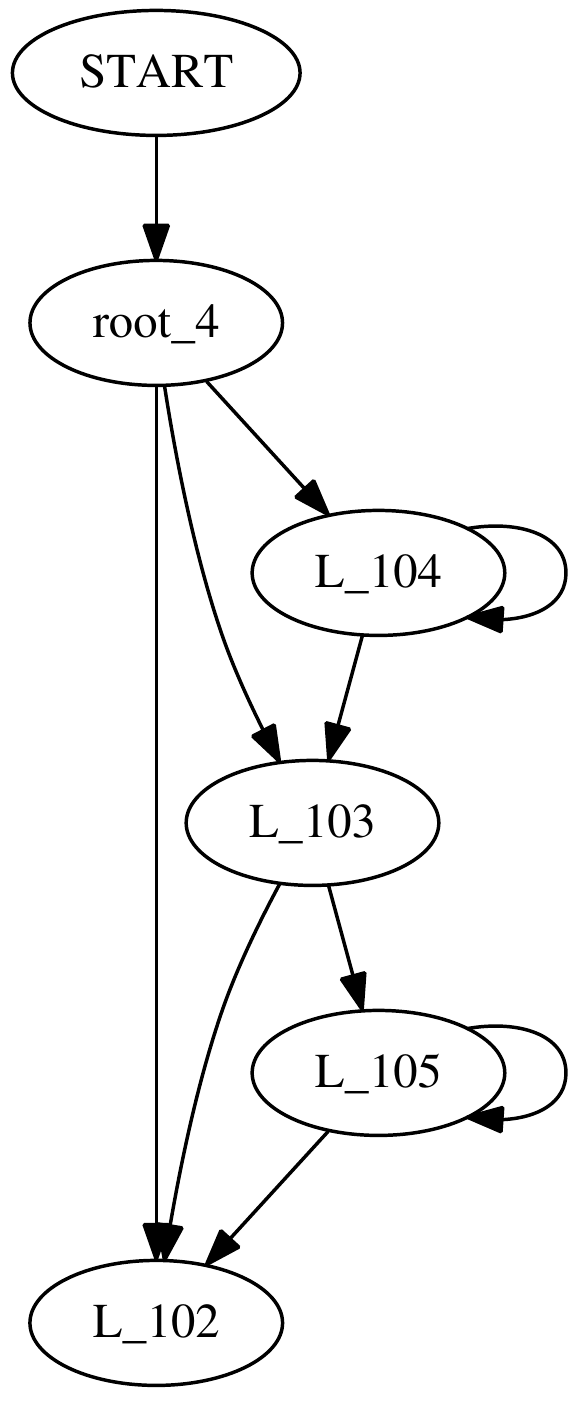} & \includegraphics[scale=.18]{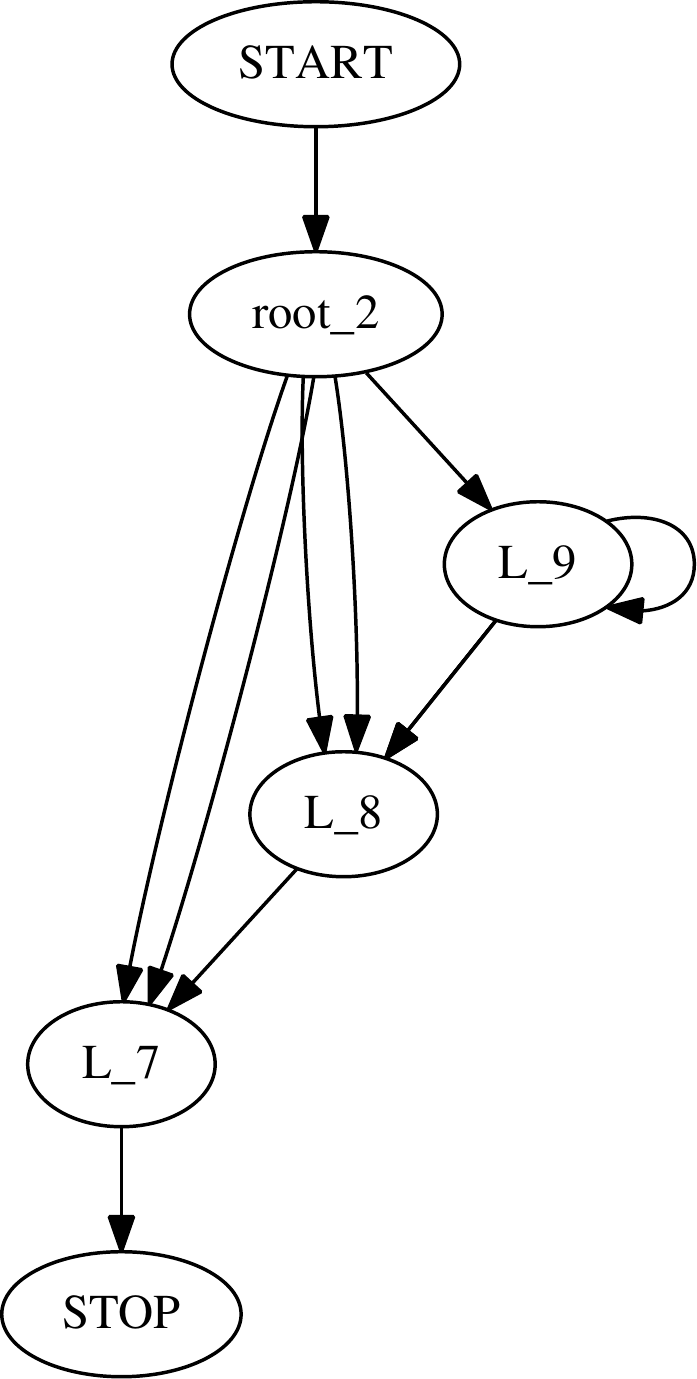} & \includegraphics[scale=.12]{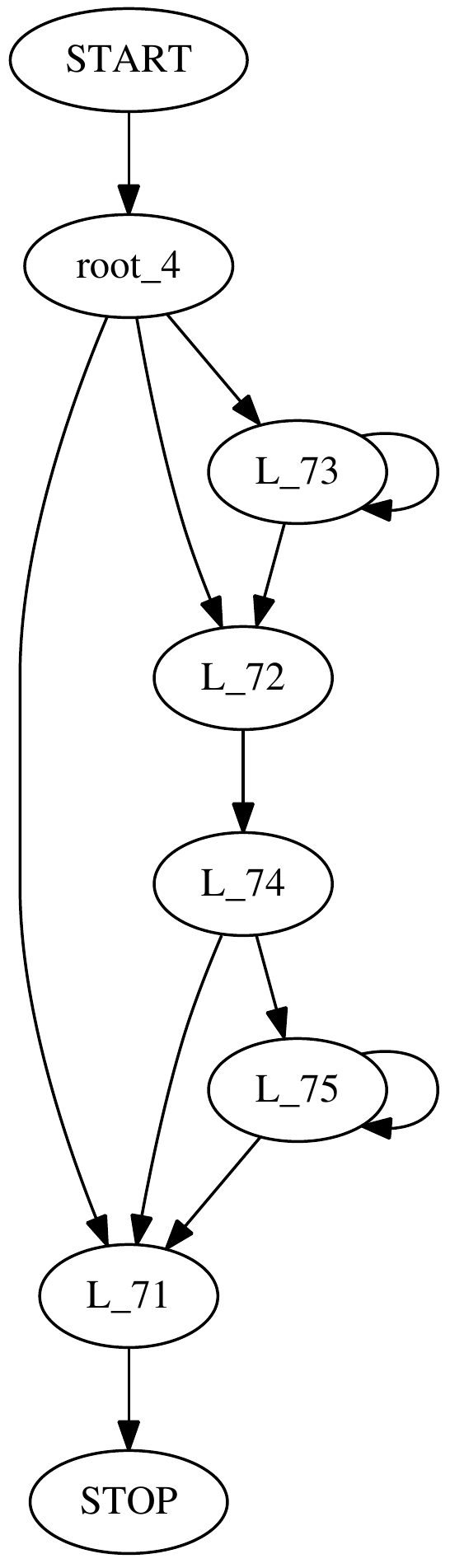} & \includegraphics[scale=.12]{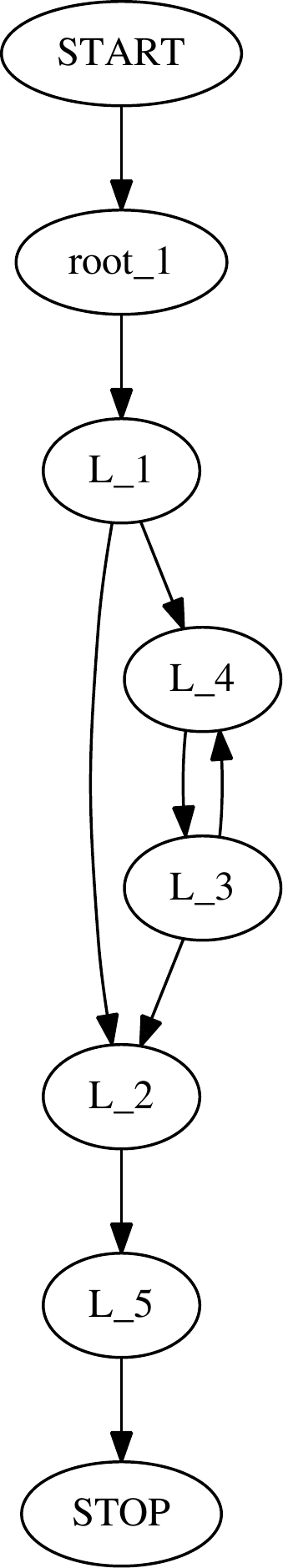} & \includegraphics[scale=.15]{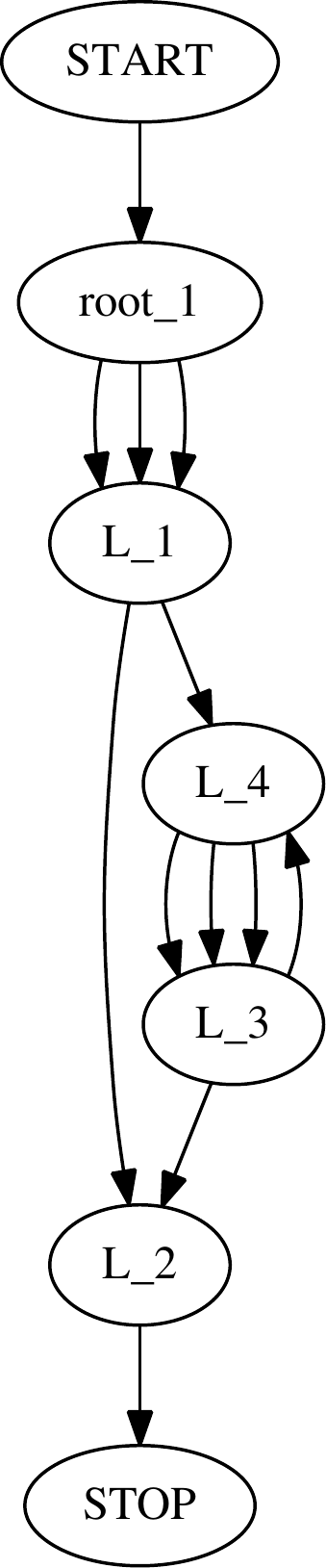} & \includegraphics[scale=.15]{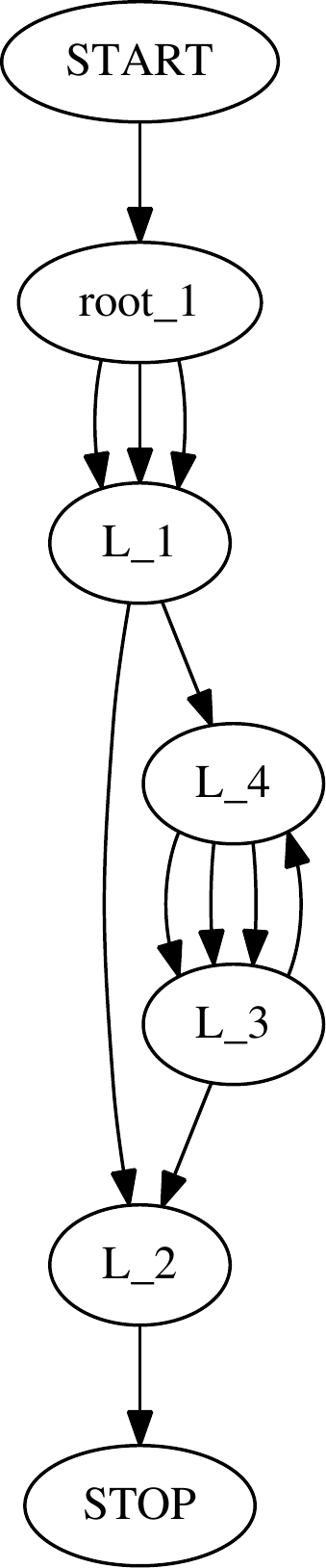} \\ \hline
 \multicolumn{3}{|p{0.3\columnwidth}|}{\centering\tabhead{ \begin{scriptsize} Hotspot \texttt{calc\_temp} \end{scriptsize}}} & \multicolumn{3}{|p{0.3\columnwidth}|}{\centering\tabhead{\begin{scriptsize} Particlefilter \texttt{sum\_kernel} \end{scriptsize}}} & \multicolumn{3}{|p{0.3\columnwidth}|}{\centering\tabhead{ \begin{scriptsize}Pathfinder \texttt{dynproc\_kernel} \end{scriptsize}}}\\ \hline
\end{tabular}
\caption{Control flow graphs generated for each CUDA kernel, comparing architecture families (Kepler, Maxwell, Pascal).}
\label{tab:cfg}
\end{figure}


\section{Background}
\label{sec:background}

Our \texttt{CUDAflow} approach shown
in Figure~\ref{fig:flow} works in association with
the current \texttt{nvcc} toolchain.
Control flow graphs are constructed from static code analysis and program
execution statistics are gathered dynamically through program counter
sampling.  This measurement collects counts of executed instructions and
corresponding source code locations, among other information.  In this way,
the \texttt{CUDAflow} methodology provides a more accurate characterization
of the application kernel, versus hardware performance counters alone,
which lack the ability to correlate performance with source line
information and are prone to miscounting events
\cite{lim2014computationally}.
In particular, it gives a way to understand the control flow behavior during execution.

\subsubsection{Kernel Control Flow Graphs}
\label{sec:cfg}
One of the more complex parameters used to characterize SIMD thread
divergence is by using a control flow graph (CFG) representation of the
computation.
A CFG is constructed for each GPU kernel computation in program order and can be represented
as a directed acyclic graph $G = (N,E,s)$, where $(N,E)$ is a finite directed
graph, and a path exists from the $START$ node $s \in N$ to every other
node.  A unique $STOP$ node is also assumed in the CFG.  A node in the
graph represents a basic block (a straight line of code without jumps or
jump targets), whereas directed edges represent jumps in the control flow.

Each basic block region is incremented with the number of times the node is
visited.  Upon sampling the program counter, the PC address is referenced
internally to determine to which basic block region the instruction
corresponds to.
{
\begin{lstlisting}[language=c,basicstyle=\scriptsize]
.L_41:
   /*04a0*/ DSETP.LE.AND P0,PT,|R6|,+INF,PT;
   /*04a8*/ @P0 BRA `(.L_43);
   /*04b0*/ LOP32I.OR R5, R7, 0x80000;
   /*04b8*/ MOV R4, R6;

   /*04c8*/ BRA `(.L_42);
\end{lstlisting}
\label{fig:sass}        
} 

The SASS assembly code illustrates how a control flow graph is constructed.
Each basic block is labeled in the left margin (e.g. ``.L\_41''), with predication and 
branch instructions representing edges that lead to corresponding block
regions (e.g. ``.L\_43,'' ``.L\_42'').  The PC offsets are listed in
hexadecimal between the comments syntax ($\mathit{\\/\ast \ast\\/}$). In
other words, ``.L\_41'' represents a node $n_i$, with ``.L\_43,'' and
``.L\_42'' as its children.


Example control flow graphs for selected SHOC (top)
\cite{danalis2010scalable} and Rodinia (bottom) \cite{che2009rodinia} GPU
benchmarks are displayed in Figure~\ref{tab:cfg}.  Different GPU
architecture types will result in the \texttt{nvcc} compiler producing
different code and possibly control flow, as seen in the CFGs from Figure~\ref{tab:cfg} for Kepler, Maxwell and Pascal architectures.  Section~\ref{sec:experiments} discusses the
differences in GPU architectures.  The CFG differences for each architecture are due in part to
the architecture layout of the GPU and its compute capability (NVIDIA virtual architecture).  The Maxwell generally uses fewer nodes for its CFGs, as evident in \texttt{kernel\_warp}.  Our approach can expose these important architecture-specific effects on the CFGs.   Also, note that similarities in structure exist with several CFGs, including \texttt{csr\_scalar} and \texttt{sum\_kernel}.  Part of the goal of this research is to predict the required resources for the application by inferring performance through CFG subgraph matching, with the subgraphs serving as building blocks for more nested and complex GPU kernels.  For this purpose, we introduce several metrics that build on this CFG representation.

\subsubsection{Transition probability}

Transition probabilities represent frequencies of an edge to a vertex, or
branches to code regions, which describes the application in a way that
gets misconstrued in a flat profile.  A stochastic matrix could also
facilitate in eliminating dead code, where states with $0$ transition
probabilities represent node regions that will never be visited.  Kernels
employing structures like loops and control flow increase the complexity
analysis, and knowledge of transition probabilities of kernels could help during code
generation.

A canonical adjacency matrix $M$ represents a graph $G$ such that every
diagonal entry of $M$ is filled with the label of the corresponding node
and every off-diagonal entry is filled with the label of the corresponding
edge, or zero if no edge exists \cite{yan2002gspan}.  The adjacency matrix
describes the transition from $N_{i}$ to $N_{j}$.  If the probability of moving from $i$ to $j$ in one time step is $Pr(j|i)=m_{i,j}$, the adjacency
matrix is given by $m_{i,j}$ as the $i^{th}$ row and the $j^{th}$ column element.
Since the total transition probability from a state $i$ to all other states
must be 1, this matrix is a right stochastic matrix, so that $\sum_{j}
P_{i,j}=1$.


\begin{figure} 
\centering
\begin{align*} 
\begin{split}
  \kbordermatrix{& R_{1} & L_{1} & L_{4} & L_{3} & L_{2} & L_{5} \\
      & .21 & - & - & - & - & -\\
      & 0 & .04 & - & - & - & -\\
      & 0 & .04 & .38 & - & - & -\\
      & 0 & 0 & 0 & .08 & - &  -\\
      & 0 & 0 & 0 & 0 & .21 & - \\
      & 0 & 0 & 0 & 0 & .02 & 0 } 
    \kbordermatrix{ & R_{1} & L_{3} & L_{2} & L_{1} \\
      & .30 & - & - & - \\
      & 0 & .51 & - & - \\
      & 0 & 0 & 0 & - \\
      & 0 & 0 & 0 & .21} \qquad          
      \\                    
\end{split}
\end{align*}
\caption{Transition probability matrices for Pathfinder (\texttt{dynproc\_kernel}) application, comparing Kepler (left) and Maxwell (right) versions.}
\label{fig:tm_spmv}
\end{figure}

Figure~\ref{fig:tm_spmv} illustrates transition probability matrices for a
kernel from the Pathfinder application (Tab.~\ref{tab:cfg}, bottom-rt.),
comparing Kepler (left) and Maxwell (right) versions.  Note that the Pascal version was the same as Maxwell, as evident in Fig.~\ref{tab:gpu}, lower-right, and was left out intentionally. The entries of the transition probability matrix were calculated by normalizing over the total number of observations for each observed node transition $i$ to $j$. Although the matrices differ in size, observe that a majority of the transitions take place in the upper-left triangle, with a few transitions in the bottom-right, for all
matrices.  The task is to match graphs of arbitrary sizes based on its
transition probability matrix and instruction operations executed, among
other information.

\subsubsection{Hybrid Static and Dynamic Analysis}
\label{sec:implementation}

We statically collect instruction mixes and source code locations from
generated code and map the instruction mixes to the source locator activity
as the program is being run \cite{lim2015identifying}.  The static analysis
of CUDA binaries produces an objdump file, which provides assembly
information, including instructions, program counter offsets, and source
line information.  The CFG structure is stored in iGraph format
\cite{csardi2006igraph}.  We attribute the static analysis from the objdump
file to the profiles collected from the source code activity to provide
runtime characterization of the GPU as it is being executed on the
architecture.  This mapping of static and dynamic profiles provides a rich
understanding of the behavior of the kernel application with respect to the
underlying architecture.

\section{Methodology}
\label{sec:methodology}

Based on the kernel CFG and transition probability analysis, the core of
the \texttt{CUDAflow} methodology focuses on the problem of subgraph
matching.  In order to perform subgraph matching, we first scale the
matrices to the same size by taking for graphs $G_1$ and $G_2$ the maximal
proper submatrix, constructed by $\mathcal{B}(G_{i}) =
\text{max}(|V_1|,|V_2|)$ for a given $G_i = \text{min}(|V_1|,|V_2|)$ using
spline interpolation.  The similarities in the shapes of the control flow
graphs, the variants generated for each GPU (Table~\ref{tab:cfg}) and the activity regions in the transition probability matrices (Fig.~\ref{fig:tm_spmv}) provided motivation for this approach.  In our case, the dense hotspots in the transition matrix should align with their
counterparts if the matrices are similar enough.

\subsection{Bilinear Interpolation}

To scale the transition matrix before performing the pairwise comparison,
we employ a spline interpolation procedure.  Spline interpolation is
general form of linear interpolation for functions of $n$-order polynomial,
such as bilinear and cubic.  For instance, a spline on a two-order
polynomial performs bilinear interpolation on a rectilinear 2D grid
(e.g. $x$ and $y$) \cite{gonzales1993digital}.  The idea is to perform
linear interpolation in both the vertical and horizontal directions.
Interpolation works by using known data to estimate values at unknown
points.  Refer to~\cite{gonzales1993digital} for the derivation of bilinear interpolation.

\begin{table}
\caption{Distance measures considered in this paper.}
\centering 
\begin{tabular}{|c|c|c|} \hline
\ Abbrev & Name & Result  \\ \hline \hline
\ Euc & Euclidean & $\sqrt{\sum_{i=1}^n | x_i - y_i |^2} $ \\ 
\ Iso & IsoRank & $(\mathbf{I} - \alpha \mathbf{Q} \times \mathbf{P}) \mathbf{x}$  \\ 
\ Man & Manhattan & $\sum_{i=1}^n | x_i - y_i |$  \\ 
\ Min & Minkowski & $\sqrt[p]{ \sum_{i=1}^n | x_i - y_i |^p }$  \\ 
\ Jac & Jaccard & $\frac{\sum_{i=1}^n (x_i - y_i)^2}{\sum_{i=1}^n x_i^2 + \sum_{i=1}^n y_i^2 - \sum_{i=1}^n x_i y_i}$  \\ 
\ Cos & Cosine & $1 - \frac{\sum_{i=1}^n x_i y_i}{\sqrt{\sum_{i=1}^n x_i^2} \sqrt{\sum_{i=1}^n y_i^2}}$  \\ 
\hline\end{tabular}
\label{tab:similarity}
\end{table}

\subsection{Pairwise Comparison}

Once the matrix is interpolated, the affinity scores ($S_1$ and $S_2$ for
graphs $G_1'$ and $G_2'$, respectively) are matched via a distance
measure, which includes the Euclidean distance, the IsoRank solution~\cite{singh2007pairwise}, Manhattan distance, Minkowski metric, Jaccard similarity, and Cosine similarity.  The distance measures considered in this work are listed in Table~\ref{tab:similarity}. By definition, $\text{sim}(G_i, G_j) = 0$ when $i = j$, with the similarity measure placing progressively higher scores for objects that are further apart.   

\section{Experimental Setup}\label{sec:experiments}
To demonstrate our \text{CUDAflow} methodology, we measured the performance of applications on several GPU architectures. 

\subsection{Execution environment}
The graphic processor units used in our experiments are listed in Table~\ref{tab:gpu}.  
The selected GPUs reflect the various architecture family generations, and performance results presented in this paper represent GPUs belonging to the same family.  For instance, we observed that the performance results from a K80 architecture and a K40 (both Kepler) were similar, and, as a result, did not include comparisons of GPU architectures within families.  Also, note the changes in architectural features across generations (global memory, MP, CUDA cores per MP), as well as ones that remain fixed (constant memory, warp size, registers per block).  For instance, while the number of multiprocessors increased in successive generations, the number of CUDA cores per MP (or streaming multiprocessors, SM) actually decreased.  Consequently, the number of CUDA cores ($\text{MP} \times \text{CUDA}_{\text{cores\_per\_mp}}$) increased in successive GPU generations.

\begin{table}
\caption{Graphical processors used in this experiment.}
\centering \scriptsize
\begin{tabular}{|r|ccc|} \hline
\ &K80&M40&P100\\ \hline \hline
\ CUDA capability & 3.5 & 5.2 & 6.0 \\ 
\ Global memory (MB) & 11520 & 12288 & 16276 \\ 
\ Multiprocessors (MP) & 13 & 24 & 56 \\ 
\ CUDA cores per MP & 192 & 128 & 64 \\ 
\ CUDA cores & 2496 & 3072 & 3584 \\ 
\ GPU clock rate (MHz) & 824 & 1140 & 405 \\ 
\ Memory clock rate (MHz) & 2505 & 5000 & 715 \\ 
L2 cache size (MB) & 1.572 & 3.146 & 4.194 \\ 
Constant memory (bytes) & 65536 & 65536 & 65536 \\ 
Shared mem blk (bytes) & 49152 & 49152 & 49152 \\ 
Registers per block & 65536 & 65536 & 65536 \\ 
Warp size & 32 & 32 & 32 \\ 
Max threads per MP & 2048 & 2048 & 2048 \\ 
Max threads per block & 1024 & 1024 & 1024 \\ 
CPU (Intel) & Haswell & Ivy Bridge & Haswell \\ \hline\hline
Architecture family & Kepler & Maxwell & Pascal \\
\hline\end{tabular}
\label{tab:gpu}
\end{table}

\subsection{Applications}
Rodinia and SHOC application suite are a class of GPU applications that cover a wide range of computational patterns typically seen in parallel computing.  Table~\ref{tab:apps} describes the applications used in this experiment along with source code statistics, including the number of kernel functions, the number of associated files and the total lines of code.

\subsubsection{Rodinia}
Rodinia is a benchmark suite for heterogeneous computing 
which includes applications and kernels that target multi-core CPU and GPU platforms \cite{che2009rodinia}. Rodinia covers a wide range of parallel communication patterns, synchronization techniques, and power consumption, and has led to architectural insights such as memory-bandwidth limitations and the consequent importance of data layout.

\subsubsection{SHOC Benchmark Suite}

The Scalable HeterOgeneous Computing (SHOC) application suite is a collection of benchmark programs testing the performance and stability of systems using computing devices with non-traditional architectures for general purpose computing \cite{danalis2010scalable}. SHOC provides implementations for CUDA, OpenCL, and Intel MIC, and supports both sequential and MPI-parallel execution.

\begin{table}[thb]
    \caption{Description of SHOC (top) and Rodinia (bottom) benchmarks studied.}
  \centering \scriptsize
  \begin{tabular}{|c|r|c|c|c|p{6.5cm}|}
	\hline
	\multicolumn{1}{|p{0.02\columnwidth}|}{\centering\tabhead{}} &
	\multicolumn{1}{|p{0.08\columnwidth}|}{\centering\tabhead{\begin{scriptsize}Name\end{scriptsize}}} &
	\multicolumn{1}{|p{0.05\columnwidth}|}{\centering\tabhead{\begin{scriptsize}Ker\end{scriptsize}}} &
   \multicolumn{1}{|p{0.05\columnwidth}|}{\centering\tabhead{\begin{scriptsize}File\end{scriptsize}}} &	
	\multicolumn{1}{|p{0.05\columnwidth}|}{\centering\tabhead{\begin{scriptsize}Ln\end{scriptsize}}} &
		\multicolumn{1}{|p{0.09\columnwidth}|}{\centering\tabhead{\begin{scriptsize}Description\end{scriptsize}}}\\
	\hline\hline
	\parbox[t]{1mm}{\multirow{10}{*}{\rotatebox[origin=c]{90}{SHOC}}} & \text{FFT} & 9 & 4 & 970 & Forward and reverse 1D fast Fourier transform. \\
    & \text{MD}  & 2 & 2 &  717 & Compute the Lennard-Jones potential from molecular dynamics.\\
   & \text{MD5Hash} & 1 & 1 & 720 & Computate many small MD5 digests, heavily dependent on bitwise operations. \\
	& \text{Reduction} & 2 & 5 & 785 & Reduction operation on an array of single or double precision  floating point values. \\
	& \text{Scan} & 6 & 6 & 1035 & Scan (parallel prefix sum) on an array of single or double precision floating point values. \\
& 	\text{SPMV}  & 8 & 2 & 830 & Sparse matrix-vector multiplcation.\\	
&	\text{Stencil2D}  & 2 & 12 & 1487 & A 9-point stencil operation applied to a 2D dataset.\\
	\hline
	\hline
\parbox[t]{1mm}{\multirow{17}{*}{\rotatebox[origin=c]{90}{Rodinia}}} &	\text{Backprop}&  2 & 7 & 945 & Trains weights of connecting nodes on a layered neural network.\\
&	\text{BFS} & 2 & 3 & 971 & Breadth-first search, a common graph traversal. \\
&    \text{Gaussian} & 2 & 1 & 1564 & Gaussian elimination for a system of linear equations. \\
&	\text{Heartwall} & 1 & 4 & 6017 & Tracks changing shape of walls of a mouse heart over a sequence of ultrasound images. \\
&	\text{Hotspot}  & 1 & 1 & 1199 & Estimate processor temperature based on floor plan and simulated power measurements. \\
 &   \text{Nearest Neighbor}  & 1 & 2 & 385 & Finds k-nearest neighbors from unstructured data set using Euclidean distance. \\
&    \text{Needleman-Wunsch} & 2 & 3 & 1878 & Global optimization method for DNA sequence alignment.  \\
&	\text{Particle Filter} & 4 & 2 & 7211 & Estimate location of target object given noisy measurements in a Bayesian framework. \\
&	\text{Pathfinder} & 1 & 1 & 707 & Scan (parallel prefix sum) on an array of single or double precision floating point values. \\
&		\text{SRAD v1}  & 6 & 12 & 3691 & Diffusion method for ultrasonic and radar imaging applications based on PDEs.\\
	&	\text{SRAD v2}  & 2 & 3 & 2021 & ...  \\		
	\hline	
  \end{tabular}

  \label{tab:apps}
\end{table}


\section{Analysis}\label{sec:results}
To illustrate our new methodology, we analyzed the SHOC and Rodinia applications at different granularities.

\subsection{Application level}
\begin{figure}
\centering
\includegraphics[width=0.46\columnwidth]{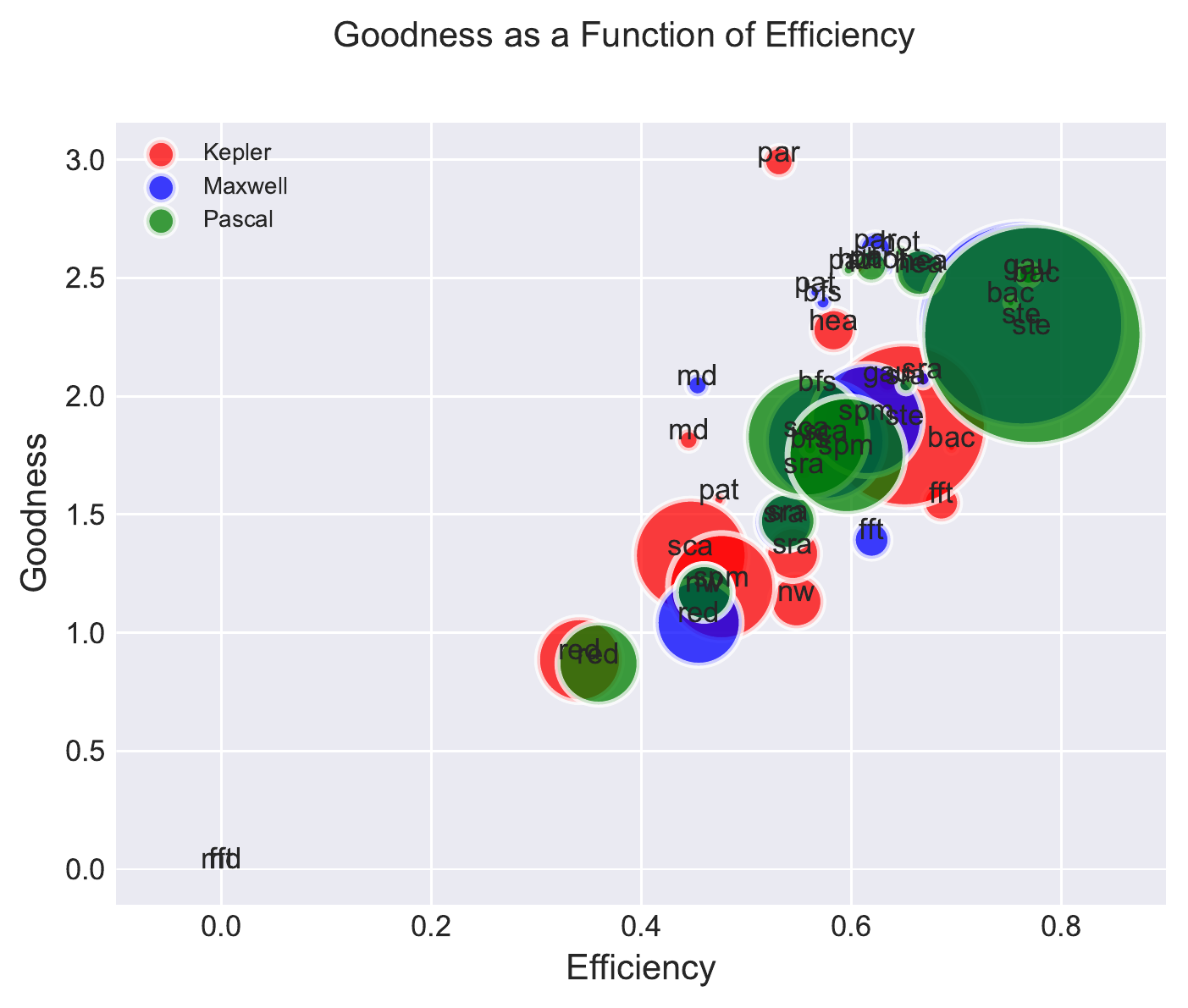}
\includegraphics[width=0.52\columnwidth]{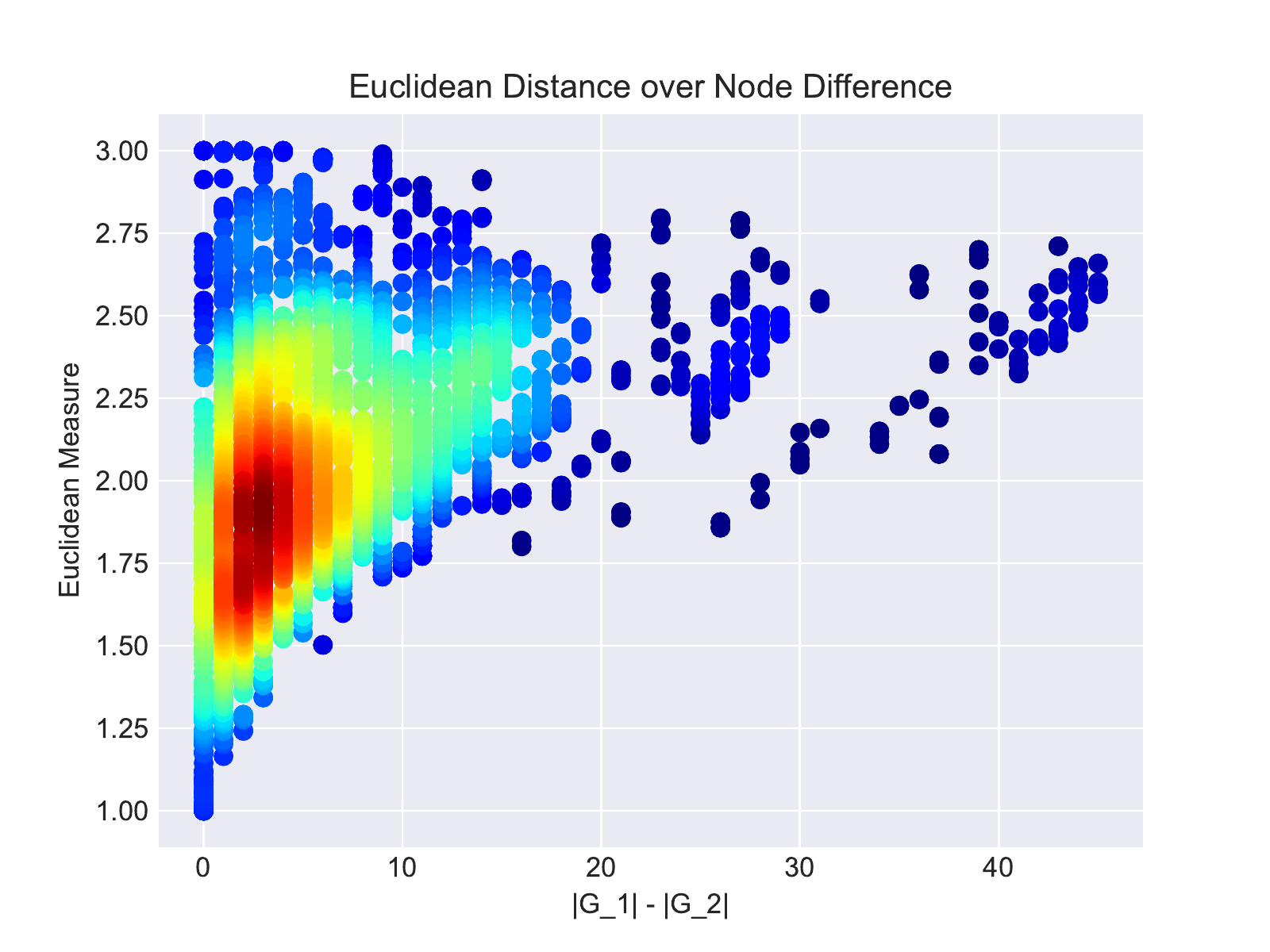}
\caption{Left:  The \emph{static} goodness metric (Eq.~\ref{eq:goodness}) is positively correlated with the \emph{dynamic}
efficiency metric (Eq.~\ref{eq:efficiency}). The color represents the architecture and the size of bubbles represents the number of operations. Right:  Differences in vertices between two graphs, as a function of Euclidean metric for all GPU kernel combinations.  Color represents intensity.}
\label{fig:bubble}
\end{figure}

Figure~\ref{fig:bubble} projects goodness as a function of efficiency, which displays the similarities and differences of the benchmark applications.  The size of bubble represents the number of operations executed, whereas the shade represents the GPU type.  
Efficiency describes how gainfully employed the GPU floating-point units remained, or FLOPs per second:

\begin{eqnarray}
\mathit{efficiency} &=&\frac{\mathit{op_{fp}}\mathit{+op_{int}}+\mathit{op_{simd}}+ \mathit{op_{conv}}}{\mathit{time_{exec}}}\cdot \mathit{calls_n}
\label{eq:efficiency}
\end{eqnarray}
The \emph{goodness} metric describes the intensity of the floating-point and memory operation arithmetic intensity:
\begin{eqnarray}
\mathit{goodness} &=&\sum_{j\in J} \mathit{op_j}\cdot \mathit{calls_n}
\label{eq:goodness}
\end{eqnarray}
Note that efficiency is measured via runtime, whereas goodness is measured statically.  Figure~\ref{fig:bubble} (left) shows a positive correlation between the two measures, where the efficiency of an application increases along with its goodness.  Static metrics, such as \textit{goodness}, can be used to derive dynamic behavior of an application.  This figure also demonstrates that  merely counting the number of executed operations is not sufficient to characterize applications because operation counts do not fully reveal control flow, which is a source of bottlenecks in large-scale programs.

\begin{figure}
\centering
\includegraphics[width=0.9\columnwidth]{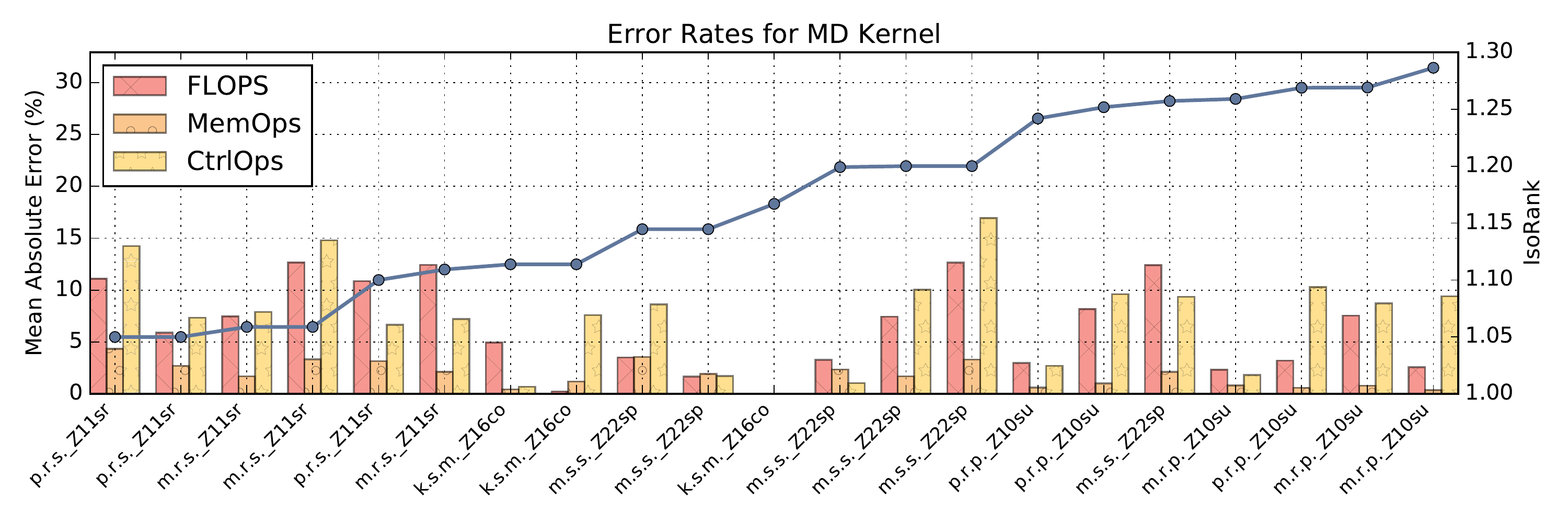}\\
\includegraphics[width=0.5\columnwidth]{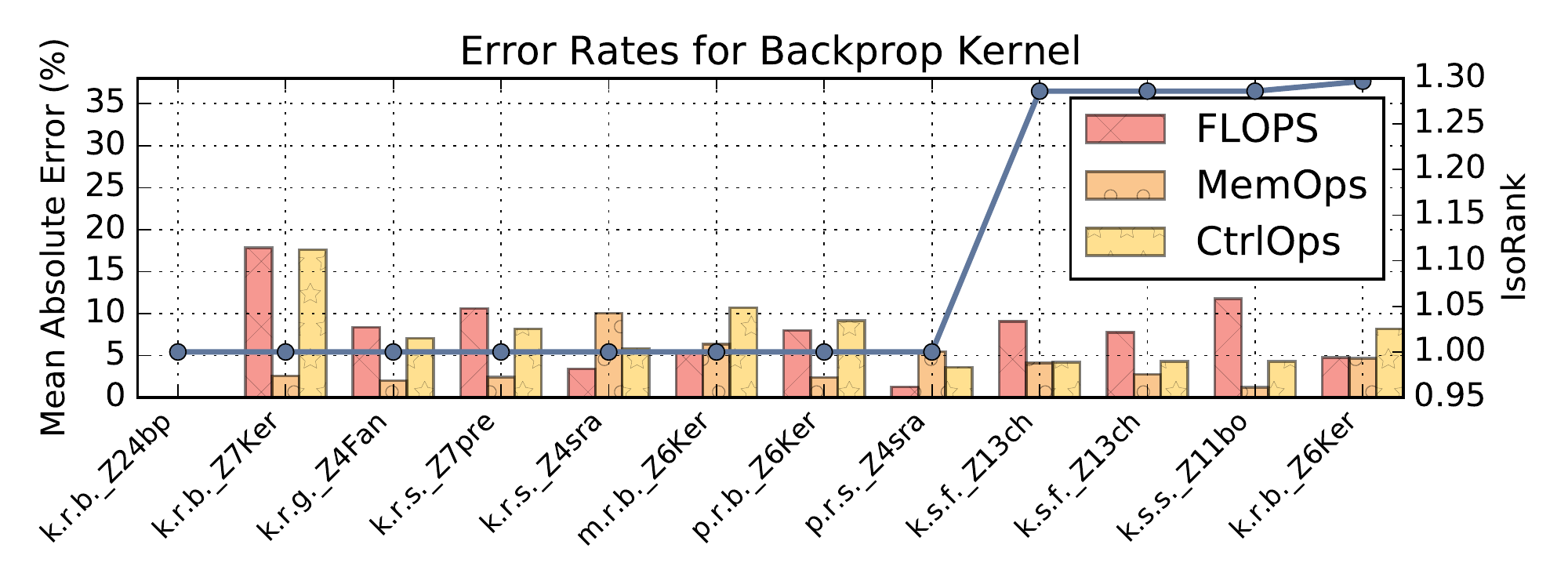}
\includegraphics[width=0.4\columnwidth]{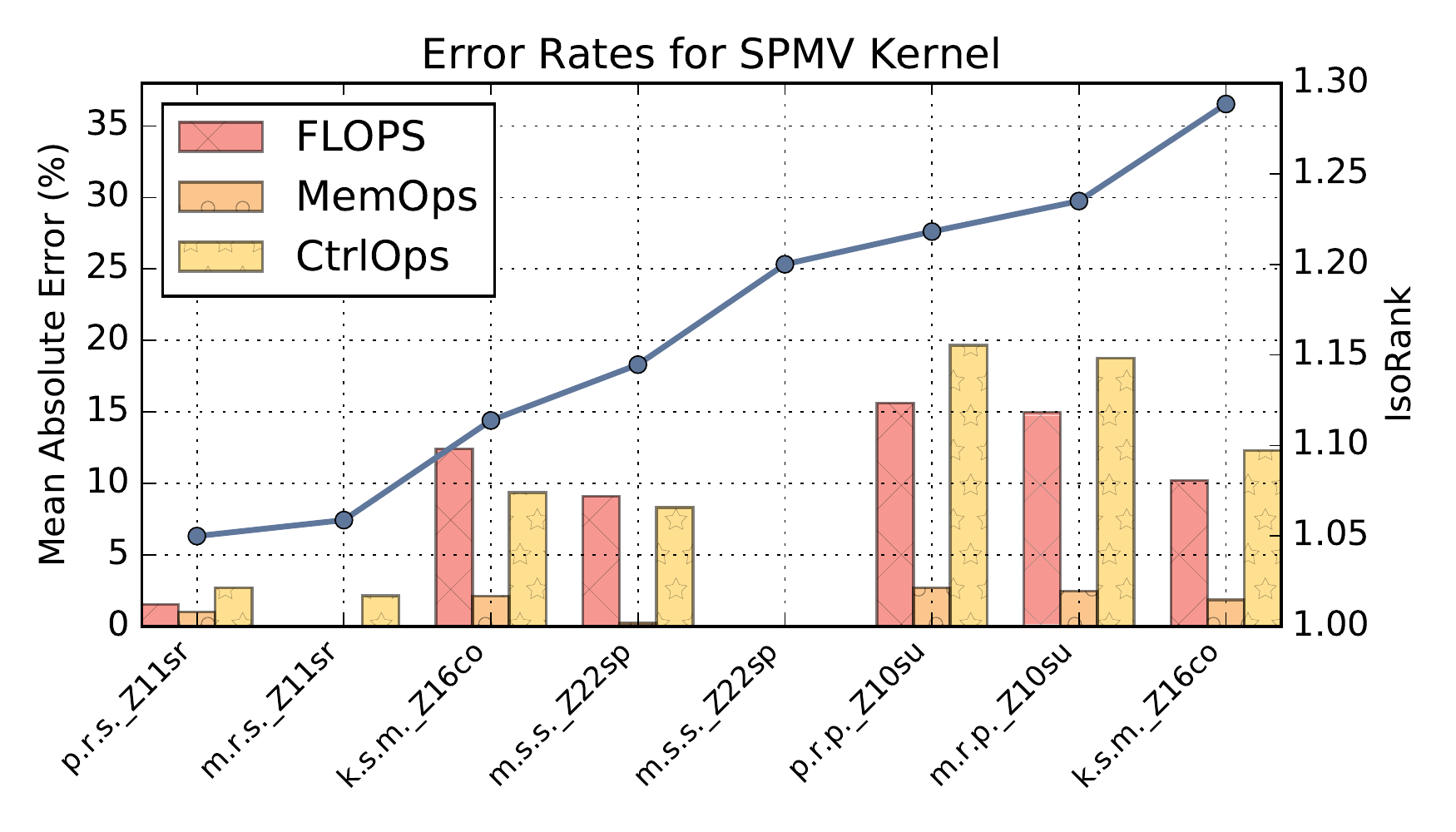}
\caption{Error rates when estimating instruction mixes statically from runtime observations for selected matched kernels (x-axis), with IsoRank scores near 1.30.}
\label{fig:MD}
\end{figure}

\subsection{CFG subgraph matching}

%

\begin{figure*}
\centering
\includegraphics[width=0.9\columnwidth]{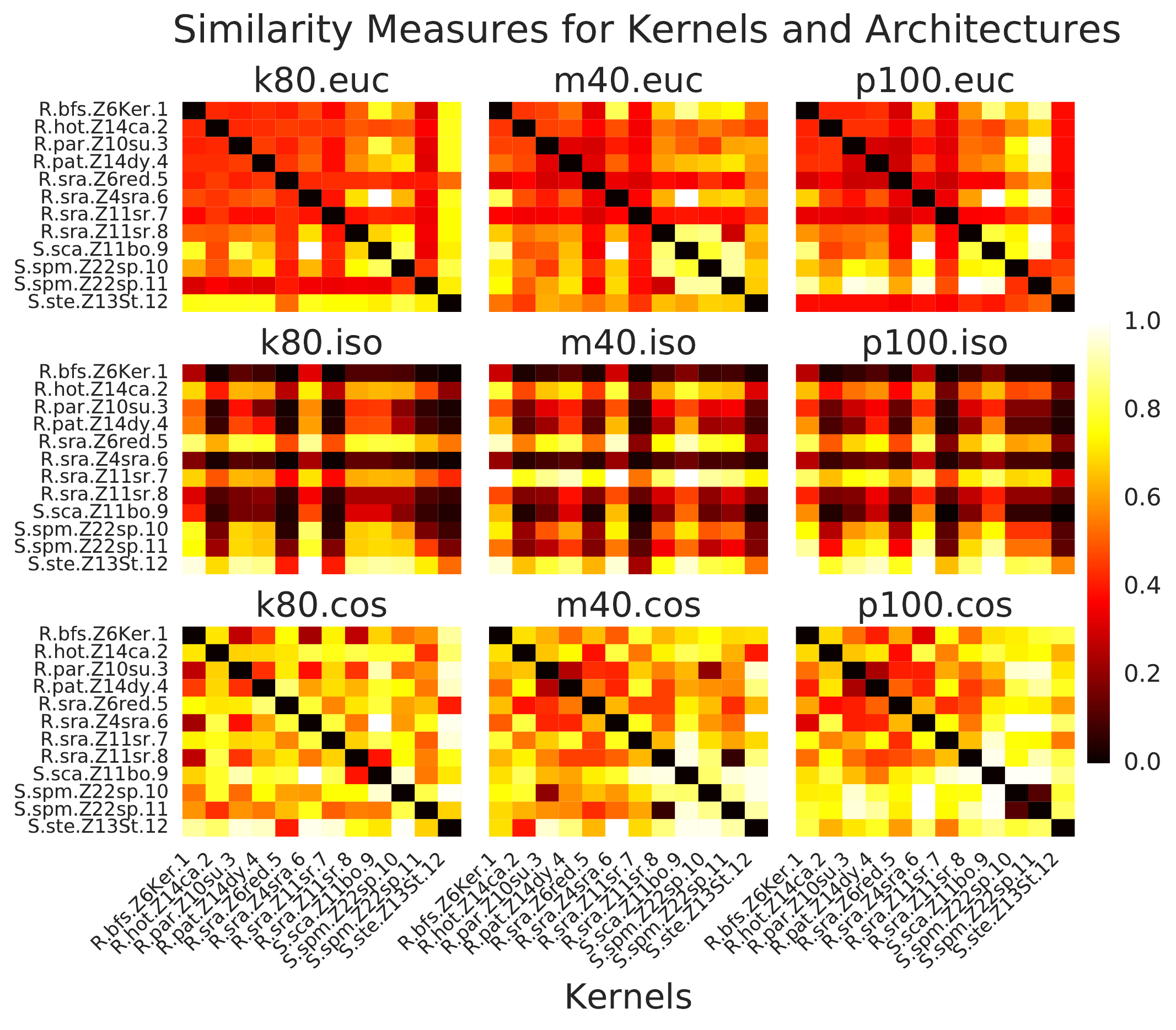}
\caption{Similarity measures for Euclidean, IsoRank and Cosine distances for 12 arbitarily selected kernels.}
\label{fig:heatmap1}
\end{figure*}

\subsubsection{Distribution of Matched Pairs}

Figure~\ref{fig:bubble} (right) projects the distribution of differences in vertices $|V|$ for all 162 CFG kernel pairs (Table~\ref{tab:apps}, 2nd col. + 3 GPUs) as a function of the Euclidean measure (application, architecture, kernel), with shade representing the frequency of the score.  Note that most matched CFGs had a similarity score of 1.5 to 2.2 and had size differences under 10 vertices.  Figure~\ref{fig:bubble} (right) also shows that as the differences in vertices increase, similarity matching becomes degraded due to the loss of quality when interpolating missing information, which is expected.  Another observation is that strong similarity results when node differences of the matched kernel pairs were at a minimum, between 0 and 8 nodes.  

\subsubsection{Error Rates from Instruction Mixes}
Here, we wanted to see how far off our instruction mix estimations were from our matched subgraphs.  Figure~\ref{fig:MD} displays instruction mix estimation error rates, calculated using mean squared error, for MD, Backprop, and SPMV kernels as a function of matched kernels (x-axis) with IsoRank scores between 1.00 to 1.30.  Naming convention for each kernel is as follows:  \textit{$\langle$gpu\_arch.suite.app.kernel$\rangle$}.  In general, \texttt{CUDAflow} is able to provide subgraph matching for arbitrary kernels through the IsoRank score in addition to instruction mixes within a 8\%  margin of error.  Note that since relative dynamic performance is being estimated from static information, the error rates will always be high.  

\begin{figure*}
\centering
\includegraphics[width=0.9\columnwidth]{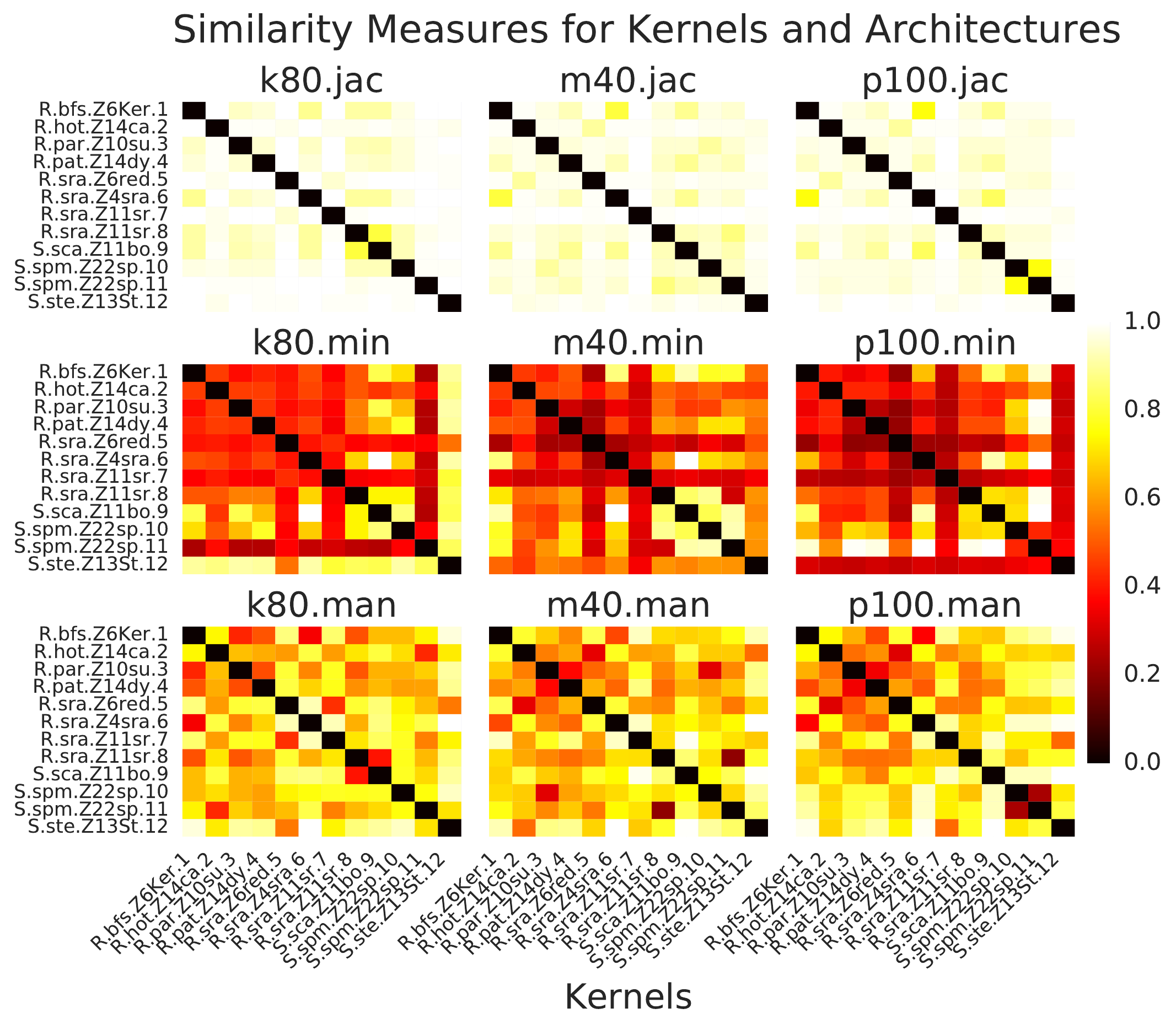}
\caption{Similarity measures for Jaccard, Minkowski and Manhattan distances for 12 arbitarily selected kernels.}
\label{fig:heatmap2}
\end{figure*}

\subsubsection{Pairwise Matching of Kernels}

Figure~\ref{fig:heatmap1} shows pairwise comparisons for 12 arbitrary selected kernels, comparing Euclidean (top), IsoRank (middle), and Cosine distance (bottom) matching strategies, and GPU architectures (rows).  Figure~\ref{fig:heatmap2} shows comparisons for the Jaccard measure, Minkowski, and Manhattan distances for the same 12 kernels.  Note that the distance scores were scaled to 0 and 1, where 0 indicates strong similarity and 1 denotes weak similarity.  In general, all similarity measures, with the exeception of IsoRank, is able to match against itself, as evident in the dark diagonal entries in the plots.  However, this demonstrates that using similarity measures in isolation alone is not sufficient for performing subgraph matching for CUDA kernels.

\begin{figure}
\centering
\includegraphics[width=0.45\columnwidth]{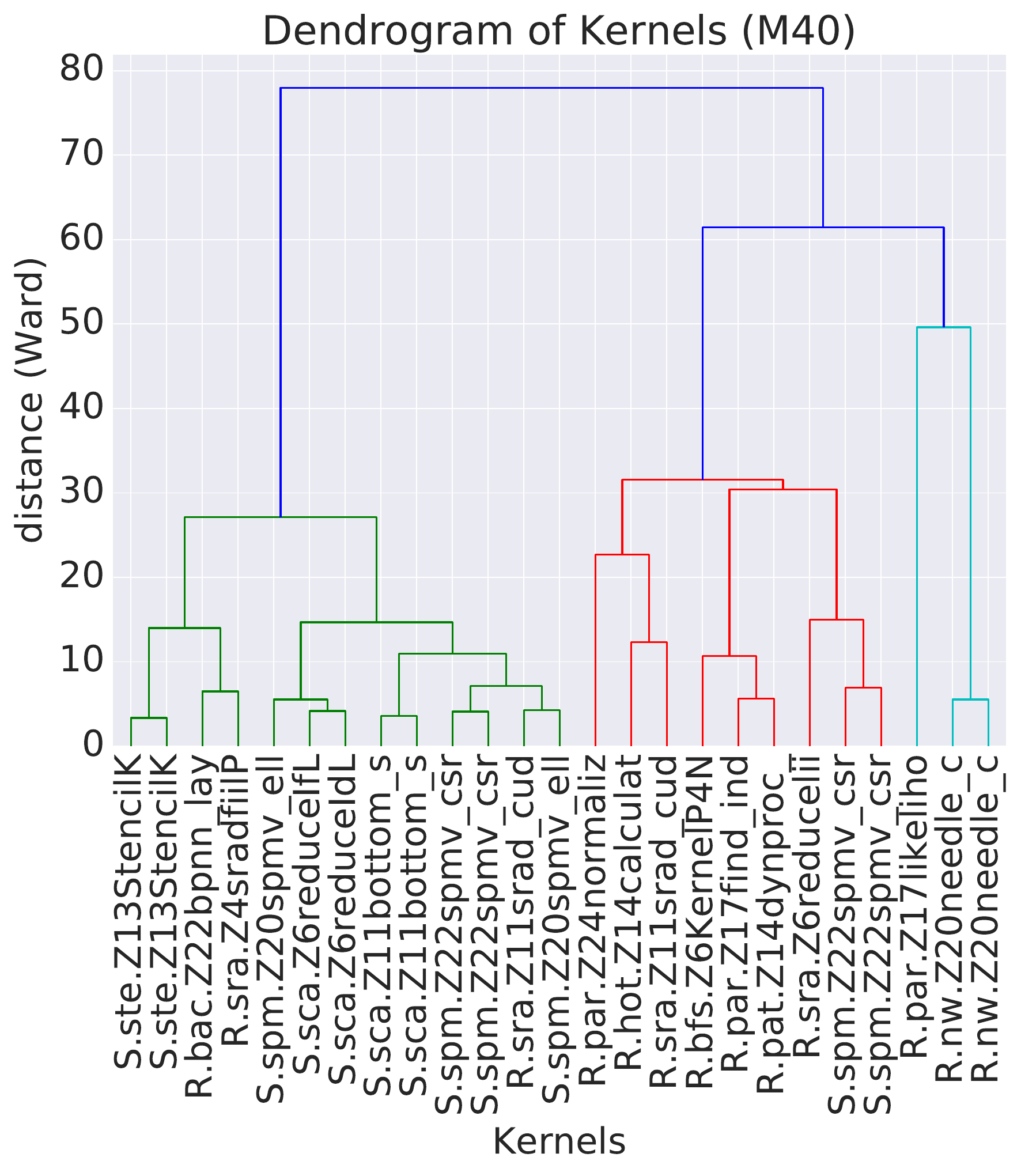}
\includegraphics[width=0.45\columnwidth]{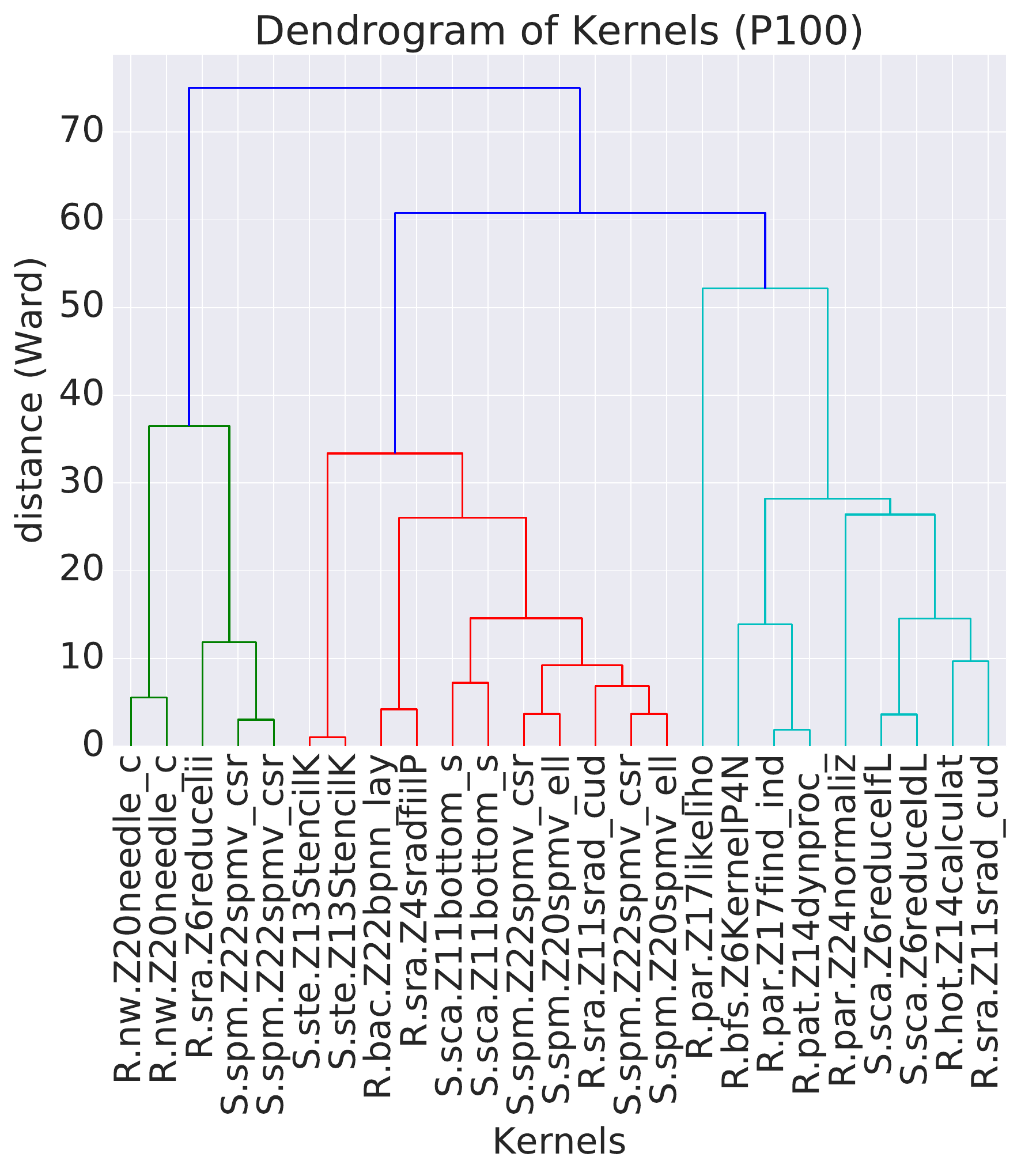}
\caption{Dendrogram of clusters for 26 kernels, comparing Maxwell (left) and Pascal (right) GPUs.}
\label{fig:hier1}
\end{figure}

\subsubsection{Clustering of Kernels}
We wanted to identify classes of kernels, based on characteristics such as instruction mixes, graph structures and distance measures.  The Ward variance minimization algorithm minimizes the total within-cluster variance by finding a pair of clusters that leads to a minimum increase in a weighted squared distances.  The initial cluster distances in Ward's minimum variance method is defined as the squared Euclidean distance between points:  $d_{ij} = d(\{ X_i \}, \{ X_j \}) = || X_i - X_j ||^2$.  Figure~\ref{fig:hier1} shows a dendrogram of clusters for 26 kernels calculated with Ward's method all matched with Rodinia Particlefilter \texttt{sum\_kernel}, comparing the Maxwell (left) and Pascal (right) GPUs, which both have 4 edges and 2 vertices in their CFGs.  \texttt{sum\_kernel} performs a scan operation and is slightly memory intensive ($\sim$26\% on GPUs).  As shown, our tool is able to categorize kernels by grouping features, such as instruction mixes, graph structures, and distance measures that show strong similarity.  This figure also demonstrates that different clusters can be formed on different GPUs for the same kernel, where the hardware architecture may result in different cluster of kernel classes.

\begin{figure}
\centering
\includegraphics[width=0.75\columnwidth]{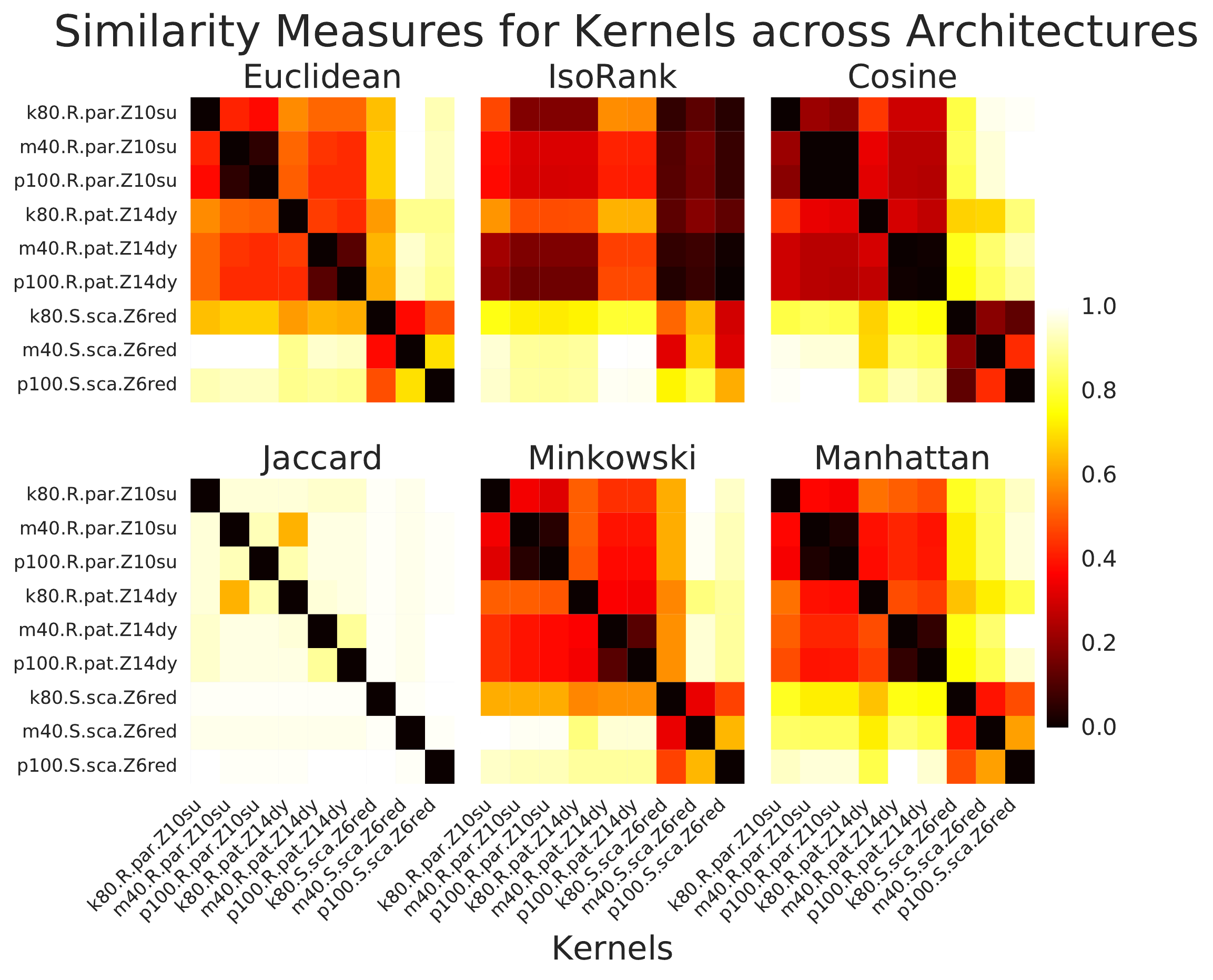}
\caption{Dendrogram of clusters for pairwise comparison for 3 kernels across 3 GPUs (9 total).}
\label{fig:heatmap3}
\end{figure}

\begin{figure}
\centering
\includegraphics[width=1\columnwidth]{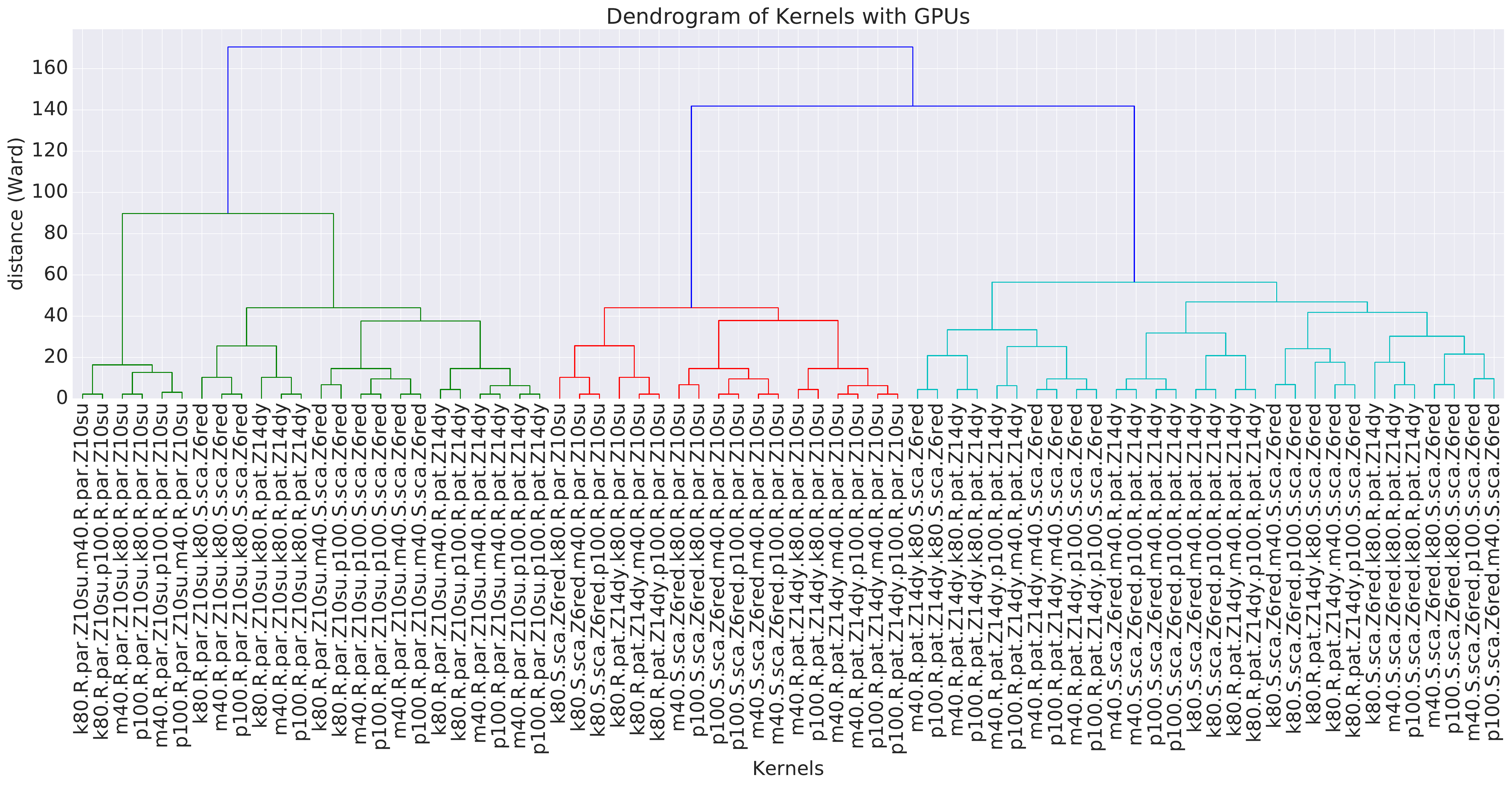}
\caption{Dendrogram of clusters for pairwise comparison for 3 kernels across 3 GPUs (27 total).}
\label{fig:hierall}
\end{figure}

Finally, we wanted to see if our technique could identify the same kernels running on a different GPU.  Figure~\ref{fig:heatmap3} shows distance measures when comparing three kernels across three GPUs, for a total of 9 comparisons, whereas Figure~\ref{fig:hierall} shows pairwise comparisons for the same three kernels across 3 GPUs, for a total of 27 comparisons (x-axis), considering pairwise comparisons in both directions (e.g. sim$(G_1, G_2)$ and sim$(G_2, G_1)$).  Figure~\ref{fig:heatmap3} displays patches of dark regions in distance measures corresponding to the same kernel when compared across different GPUs.  As shown in Figure~\ref{fig:hierall}, our tool not only is able to group the same kernel that was executed on different GPUs, as evident in the three general categories of clusters, but also kernels that exhibited similar characteristics when running on a particular architecture, such as instructions executed, graph structures, and distance measures.

\subsection{Discussion}

These metrics can be used both for guiding manual optimizations and by compilers or autotuners. For example, human optimization effort can focus on the code fragments that are ranked high by kernel impact, but low by the goodness metric.  An autotuner can also use metrics such as the goodness metric to explore the space of optimization parameters more efficiently, such as by excluding cases where we can predict a low value of the goodness metric without having to execute and time the actual generated code.  A benefit to end users (not included in paper, due to space purposes) would be providing the ability to compare an implementation against a highly optimized kernel.  By making use of subgraph matching strategy as well as instruction operations executed, \texttt{CUDAflow} is able to provide a mechanism to characterize unseen kernels.

\section{Conclusion}
\label{sec:conclusions}
We have presented \texttt{CUDAflow}, a control-flow-based methodology for analyzing the performance of CUDA applications.  We combined static binary analysis with dynamic profiling to produce a set of metrics that not only characterizes the kernel by its computation requirements (memory or compute bound), but also provides detailed insights into application performance.  Specifically, we provide an intuitive visualization and metrics display, and correlate performance hotspots with source line and file information, effectively guiding the end user to locations of interest and revealing potentially effective optimizations by identifying similarities of new implementations to known, autotuned computations through subgraph matching. We implemented this new methodology and demonstrated its capabilities on SHOC and Rodinia applications.

Future work includes incorporating memory reuse distance statistics of a kernel to characterize and help optimize the memory subsystem and compute/memory overlaps on the GPU.  In addition, we want to generate robust models that will discover optimal block and thread sizes for CUDA kernels for specific input sizes without executing the application~\cite{lim2017autotuning}.  Last, we are in the process of developing an online web portal \cite{cknowledge,sreepathi2014application} that will archive a collection of control flow graphs for all known GPU applications.  For instance, the web portal would be able to make on-the-fly comparisons across various hardware resources, as well as other GPU kernels, without burdening the end user with hardware requirements or software package installations, and will enable more feature rich capabilities when reporting performance metrics.

\bibliographystyle{splncs04}
\bibliography{typeinst}

\end{document}